\documentclass{aa}  
\usepackage{graphicx}
\usepackage{txfonts}
\usepackage{natbib}
\bibpunct{(}{)}{;}{a}{}{,} 

\newcommand{\ec}{EC~23487-2424}
\newcommand{\bpm}{BPM~31594}
\newcommand{\ross}{Ross~548}
\newcommand{\bpmb}{BPM~30551}
\newcommand{\mct}{MCT~0145-2211}
\newcommand{\he}{HE~0532-5605}
\newcommand{\lnt}{L~19-2}
\newcommand{\hs}{HS~1013+0321}
\newcommand{\wdj}{WD~J0925+0509}

\usepackage[normalem]{ulem}
\usepackage{color}

\begin{document}

   \title{ZZ~Ceti stars of the southern ecliptic hemisphere re-observed by TESS}


    \author{Zs\'ofia~Bogn\'ar\inst{1,2,3}\fnmsep\thanks{\email{bognar.zsofia@csfk.org}},
         \'Ad\'am~S\'odor\inst{1,2,3},
         Ian R. Clark\inst{4}, 
         \and Steven D. Kawaler\inst{4}}
   
   \institute{
        Konkoly Observatory, E\"otv\"os Lor\'and Research Network (ELKH), Research Centre for Astronomy and Earth Sciences, Konkoly Thege Mikl\'os \'ut 15-17, H--1121, Budapest, Hungary
        \and
        MTA CSFK Lend\"ulet Near-Field Cosmology Research Group
        \and
        MTA Centre of Excellence
        \and
        Department of Physics and Astronomy, Iowa State University, Ames, IA 50011 USA}
        
    \titlerunning{ZZ~Ceti stars re-observed by TESS}
        \authorrunning{Zs.~Bogn\'ar et al.}
        
    \date{}

 
  \abstract
   {In 2020, a publication presented the first-light results for 18 known ZZ~Ceti stars observed by the TESS space telescope during the first survey observations of the southern ecliptic hemisphere. However, in the meantime, new measurements have become available from this field, in many cases with the new, 20s ultrashort cadence mode.}
   {We investigated the similarities and differences in the pulsational behaviour of the observed stars between the two observational seasons, and searched for new pulsation modes for asteroseismology.}
   {We performed Fourier analysis of the light curves using the standard pre-whitening process, and compared the results with frequencies obtained from the earlier data. Utilising the 2018 version of the White Dwarf Evolution Code, we also performed an asteroseismic analysis of the different stars. We searched for models with seismic distances in the vicinity of the Gaia geometric distances.}
  {We detected several new possible pulsation modes of the studied pulsators. In the case of \he, we found a similar brightening phase to the one presented in the 2020 first-light paper, which means this phenomenon is recurring. Therefore, \he\ appears to be a new outbursting DAV star. We also detected a lower-amplitude brightening phase in the star \wdj. However, this case has proven to be the result of the passage of a Solar System object in the foreground. We accept asteroseismic model solutions for six stars.}
   {}


   \maketitle
%

\section{Introduction}

ZZ~Ceti or DAV stars are short-period ($P\sim100-1500$\,s), low-amplitude ($A\sim0.1\%$) variables, which means precise and short-exposure-time measurements are required both from the ground and space in order to study their pulsations. Their atmospheres are dominated by hydrogen, and they form the most populous group of pulsating white dwarf stars, lying in the 10\,500--13\,000\,K effective temperature range. Their pulsation modes are low-spherical-degree ($\ell = 1$ and $2$), and low-to-mid radial-order $g$-modes. The $\kappa-\gamma$ mechanism \citep{1981A&A...102..375D, 1982ApJ...252L..65W} in combination with the convective driving mechanism \citep{1991MNRAS.251..673B, 1999ApJ...511..904G} is responsible for the excitation of the observed pulsations.

Despite the relatively narrow instability strip of the DAV stars, they can show a large variety of pulsational properties, evolving from the hot (blue) to the cool (red) edge of the instability domain. Hotter DAVs show fewer and lower-amplitude pulsation modes than the cooler ones. Furthermore, in this latter case, above about $800\,$s in period, we usually do not detect one single peak at a given frequency in the Fourier transform of the light curve, but a large number of peaks under a broad envelope, which is reminiscent of stochastically driven oscillations. We refer the reader to \citet{2017ApJS..232...23H} for a detailed description of both the DAV instability strip and this interesting phenomenon based on observations with the Kepler space telescope. For reviews of the theoretical and observational aspects of studies of white dwarf pulsators, we also recommend the papers of \citet{2008ARA&A..46..157W}, \citet{2008PASP..120.1043F}, \citet{2010A&ARv..18..471A}, \citet{2019A&ARv..27....7C}, and \citet{2020FrASS...7...47C}.

We also mention the so-called outburst phenomena in cool DAV stars, which appears as an increase in the stellar flux of ZZ~Ceti stars close to the red edge of the instability strip. These outburst events were discovered using the measurements of the Kepler space telescope; see the papers of Keaton Bell and his collaborators: \citet{2015ApJ...809...14B, 2016ApJ...829...82B, 2017ASPC..509..303B}, and also \citet{2015ApJ...810L...5H}. Such phenomena suggest that the average brightness of the star increases relatively quickly (in about 1 hour) and by at least several per cent, and remains in this state for several hours or even longer, sometimes for even more than 1 day. After that, the stellar brightness decreases to the initial value; the outburst event repeats after several days or weeks. The duration and occurrence of these events is irregular and unpredictable. \citet{2017ASPC..509..303B} discuss a possible explanation for the outbursts: non-linear mode coupling, which can transfer energy from a driven parent mode into two daughter modes. In this case, these otherwise damped daughter modes will deposit the additional energy at the base of the convection zone, and we observe the resulting surface heating of the star. This pulsational energy-transfer mechanism could explain the observed location of the cool edge of the ZZ~Ceti instability strip, which should be much cooler according to theoretical calculations. However, \citet{2020ApJ...890...11M} raised another possibility, namely that phase shifts of 
the travelling waves reflected from the outer turning point being close to the convection zone could also be relevant in explaining the outburst phenomenon.

This paper focuses on the study of ZZ~Ceti stars observed by the Transiting Exoplanet Survey Satellite (TESS; \citealt{2015JATIS...1a4003R}). TESS was launched on 18 April 2018, and during its two-year primary mission, it provided 30 minute (long-)cadence full-frame images from almost the entire sky, and 120 second (short-)cadence observations on selected targets. The main goal of the mission is to find exoplanets at bright nearby stars with the transit method, but the time sampling of the observations also allows us to examine the pulsations of stars in the observed fields. Due to their short periods, only the short cadence mode is suitable for studying the light variations of compact pulsators. The first-light papers of the TESS Asteroseismic Science Consortium (TASC) Compact Pulsators Working Group (WG\#8), presented for example by \citet{2019A&A...632A..42B}, \citet{2019A&A...632A..90C}, and \citet{2020A&A...638A..82B}, clearly demonstrate the suitability of TESS measurements for compact pulsators. Moreover, TESS observations have significantly raised (by about 20 per cent) the number of known DAV stars \citep{2022MNRAS.511.1574R}.

Fortunately, the Extended Mission was approved for 2020--2022, with some modifications, for example in the cadence of the observations. The full-frame image cadence was reduced to 10 minutes, and a new, 20 second ultrashort cadence mode was implemented.  The latter in particular was a welcome addition given the short periods seen in compact pulsators.

This paper is the continuation of our work published in 2020 (\citealt{2020A&A...638A..82B}, hereafter P01). We detail the main goals of the present work and some points about the light curve reduction process in Sect.~\ref{sect:tess}. Section~\ref{sect:analyses} presents the results of the light-curve analyses, while Sect.~\ref{sect:seism} summarises our asteroseismic investigations. Finally, the summary and conclusions are presented in Sect.~\ref{sect:disc}.

\section{TESS observations in ultrashort cadence mode}
\label{sect:tess}

While certain pulsating white dwarfs can be used to measure evolutionary effects (cooling and contraction) by taking advantage of the very stable modes observed in their cases, other white dwarfs show short-term (on daily, weekly, or monthly timescales) amplitude and frequency variations. This latter phenomenon makes it difficult to determine the pulsation periods, because it causes the appearance of several peaks in the  Fourier transform of the light curve instead of one well-defined peak in a given frequency domain. Another difficulty is that as a result of the amplitude variations, one data set may reveal only a subset of the possible pulsation periods. However, observing the star at different epochs, we can complete the list of observed modes, which means stronger constraints for the asteroseismic modelling. 

Finding an adequate explanation for these short-term variations is a challenge for theory. As  discussed by \citet{2020ApJ...890...11M}, non-linear resonant energy transfer can be responsible for part of the short-term amplitude and frequency variations. In addition, we have to consider that the strength of the interaction with the surface convection zone of the white dwarf is different for the short- and long-period modes. In the latter case, the interaction is stronger, because longer period modes have turning points much closer to the surface than in the case of the short-period modes.  

In P01, we presented the results of the light-curve analyses of 18 previously known ZZ~Ceti stars, observed by TESS during the first survey of the southern ecliptic hemisphere. TESS returned to these fields performing further observing runs; we now have a unique opportunity to check for changes in the pulsation behaviour of these stars between the epochs. Moreover, this time, TESS observed 13 out of the 18 original targets in ultrashort cadence mode with 20\,s exposure times, which helps us to distinguish real pulsation frequencies from their Nyquist aliases. In this work, we analyse the ultrashort cadence data.

There are some differences between the targets presented in this paper and in P01. Because TESS did not return to the exact positions in the second run on the southern ecliptic hemisphere, TESS did not observe the well-known pulsator HS~0507+0434B. New observations of three not-observed-to-vary (NOV) objects, MCT~2148-2911, WD~0108-001, and KUV~03442+0719 are also missing, and EC~00497-4723 was observed only in the 120 second cadence mode. On the other hand, while \hs\ was a NOV star according to our previous analysis, the new observations revealed the presence of the pulsation frequencies reported earlier in the literature.

The following four NOV stars from P01 were also found to be non-variable based on the new TESS data: HE~0031-5525, MCT~0016-2553, HS~0235+0655, and EC~11266-2217. The journal of their new, ultrashort cadence mode observations is presented in Table~\ref{tabl:journal2}.

\begin{table*}
\centering
\caption{Journal of observations of the targets not observed to vary by the new TESS data sets. Here we list the information on the ultrashort cadence mode, because these are the data analysed in this work. Here, \textit{TIC} refers to the TESS Input Catalog identifier for the object, \textit{N} is the number of data points after cleaning the light curve, $\delta T$ is the length of the data sets including gaps, and \textit{Sect.} is the serial number of the sector(s) in which the star was observed. The start time in BJD is the time of the first data point in the data set. The \textit{CROWDSAP} keyword represents the ratio of the target flux to the total flux in the TESS aperture, while \textit{FAP} is the acronym of false alarm probability.}
\label{tabl:journal2}
\begin{tabular}{lccrrcrcr}
\hline
\hline
Object & TIC & Start time & \multicolumn{1}{c}{\textit{N}} & \multicolumn{1}{c}{$\delta T$} & \textit{G} mag & Sect. & \multicolumn{1}{c}{CROWDSAP} & \multicolumn{1}{c}{$0.1\%$\,FAP} \\
& & (BJD-2\,457\,000) & & \multicolumn{1}{c}{(d)} & & & & \multicolumn{1}{c}{(mma)}\\
\hline
EC~11266-2217 & 219442838 & 2282.177 & 90\,335 & 23.8 & 16.4 & 36 & 0.19 & 8.01\\
MCT~0016-2553 & 246821917 & 2088.244 & 79\,341 & 23.4 & 15.9 & 29 & 0.13 & 5.29\\
HE~0031-5525 & 281594636 & 2088.243 & 86\,308 & 24.2 & 15.8 & 29 & 0.60 & 4.09\\
HS~0235+0655 & 365247111 & 2447.692 & 157\,281 & 50.7 & 16.5 & 42--43 & 0.94 & 4.56\\
\hline
\end{tabular}
\end{table*}

In summary, 4 of our 13 targets observed by TESS in ultrashort cadence mode did not show significant pulsation frequencies. Therefore, we present the analyses of nine ZZ~Ceti stars in the following sections. We list these objects and their log of observations in Table~\ref{tabl:journal1}.
We note that the   NOV stars in our ZZ~Ceti sample, are relatively faint, and/or suffer from severe contamination by close objects in the large TESS pixels.

\begin{table*}
\centering
\caption{Journal of observations of the targets showing significant pulsational light variations in the new TESS data sets.}
\label{tabl:journal1}
\begin{tabular}{lccrrcrcr}
\hline
\hline
Object & TIC & Start time & \multicolumn{1}{c}{\textit{N}} & \multicolumn{1}{c}{$\delta T$} & \textit{G} mag & Sect. & \multicolumn{1}{c}{CROWDSAP} & \multicolumn{1}{c}{$0.1\%$\,FAP} \\
& & (BJD-2\,457\,000) & & \multicolumn{1}{c}{(d)} & & & & \multicolumn{1}{c}{(mma)}\\
\hline
\ross\ & 029854433 & 2115.890 & 95\,196 & 25.7 & 14.2 & 30 & 0.98 & 0.91\\
\ec\ & 033986466 & 2088.244 & 87\,877 & 24.2 & 15.4 & 29 & 0.97 & 2.60\\
\bpm\ & 101014997 & 2115.887 & 198\,676 & 54.1 & 15.1 & 30--31 & 0.90 & 1.47\\
\bpmb\ & 102048288 & 2088.243 & 79\,831 & 23.3 & 15.5 & 29 & 0.96 & 2.73\\
\mct\ & 164772507 & 2115.890 & 99\,835 & 26.0 & 15.2 & 30 & 0.96 & 2.57\\
\lnt\ & 262872628 & 2333.857 & 225\,422 & 55.9 & 13.4 & 38--39 & 0.78 & 0.44\\
\hs\ & 277747736 & 2255.806 & 273\,927 & 322.9 & 15.7 & 35,45--46 & 0.96 & 2.53\\
\wdj\ & 290653324 & 2255.885 & 267\,087 & 322.8 & 15.3 & 35,45--46 & 0.51 & 1.79\\
\he\ & 382303117 & 2088.239 & 909\,194 & 301.5 & 15.4 & 29--31,33--39 & 0.73 & 1.44\\
\hline
\end{tabular}
\end{table*}

We processed the data in a similar way to that described in P01. We downloaded the light curves from the Mikulski Archive for Space Telescopes (MAST), and extracted the PDCSAP fluxes provided by the pre-search data-conditioning pipeline \citep{2016SPIE.9913E..3EJ} from the fits files. It is important to mention that this pipeline corrects the flux for each target to account for crowding from other stars. We then corrected the light curves for long-term systematic uncertainties and outlier data points. Considering the trends on much longer timescales than the pulsations, at first we divided the time strings into segments. These segments contain gaps no longer than 0.5~d. We then separately fitted and subtracted cubic splines from each segment. We used one knot point for every 1000 points to define the splines. Finally, we removed outliers with higher than 4 sigma deviation from the mean brightness. The above corrections did not affect the frequency domain of the short-period white dwarf pulsations.

\section{Light-curve analysis}
\label{sect:analyses}

We analysed the measurements with the command-line light-curve fitting program \textsc{LCfit} developed by \'A. S\'odor \citep{2012KOTN...15....1S}. Utilising an implementation of the Levenberg-Marquardt least-squares fitting algorithm, \textsc{LCfit} is capable of linear (amplitudes and phases) and non-linear (amplitudes, phases and frequencies) least-squares fittings. The program can
also handle unequally spaced measurements, and data sets with gaps inside.

We set the detection limit to be at $0.1$ per cent false-alarm probability (FAP),  as we did in P01. This means that there is a $99.9$ per cent chance that a peak reaching this limit is not simply the result of random fluctuations due to noise. We calculated the $0.1\%$ FAP threshold as we described in P01, which refers to the work of \citet{2016A&A...585A..22Z}. Table~\ref{tabl:journal1} also lists the $0.1$ per cent FAP limits in milli-modulation amplitude (mma). These values were calculated from the average local noise in the Fourier transforms considering the $0-900\,$d$^{-1}$ (0--10\,417\,$\mu$Hz) frequency domain.   
In the following sections, we summarise our findings for the different stars, including a comparison between the light-curve analysis results from the new and previous observations presented in P01. We do not give as detailed description of the results from the former ground-based observations as we did in P01; we refer to that paper for further details.


\subsection{\ross}
\label{sec:ross}

\begin{figure*}
\centering
\includegraphics[width=\textwidth]{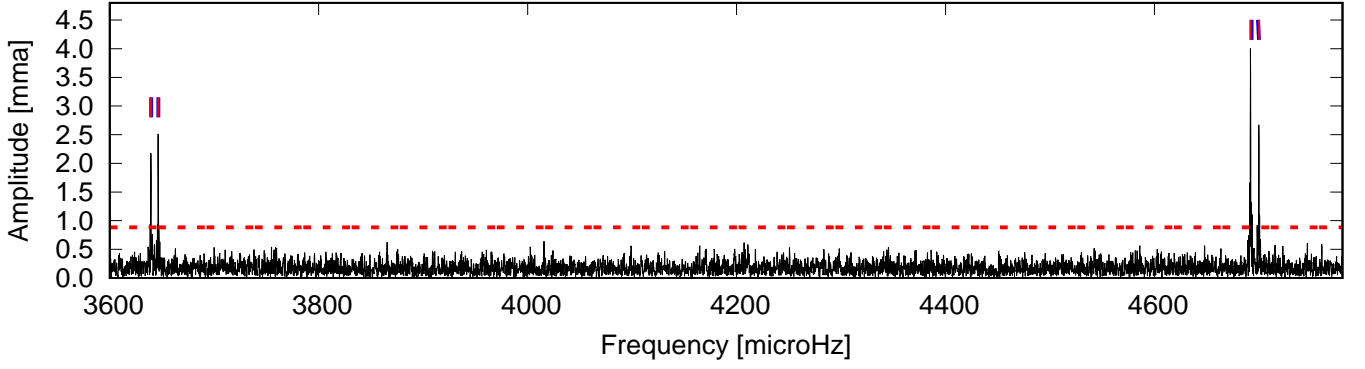}
\caption{\ross: Fourier transform of the ultrashort cadence mode light curve. We mark the frequencies detected in the previous TESS season with blue lines, while red lines denote the newly found frequencies,  and the red dashed line corresponds to the $0.1\%$ FAP significance level. We note that in this case the blue and red lines marking the individual frequencies align, because we successfully detect the previously known frequencies.}
\label{fig:ross}
\end{figure*}

The star \ross\, (TIC~029854433 $=$ ZZ~Ceti, $\alpha_{2000}=01^{\mathrm h}36^{\mathrm m}14^{\mathrm s}$, $\delta_{2000}=-11^{\mathrm d}20^{\mathrm m}33^{\mathrm s}$) is the namesake of the ZZ~Ceti or DAV class of pulsating white dwarf stars.
TESS observed \ross\ in sector 30 with the ultrashort cadence 20~s mode. Unlike in the former 120~s TESS observations, there is no Nyquist-alias problem in this new data set, and therefore direct determination of the frequencies is feasible. We  clearly detect the four highest-amplitude frequencies, but we do not find further modes reported by \citet{2015ApJ...815...56G}. Figure~\ref{fig:ross} shows the Fourier transform (FT) of the ultrashort cadence TESS light curve. In Table~\ref{tabl:ross}, we summarise the findings from both the new TESS observations and the amplitudes resulting from ground-based measurements. We also list the predicted TESS amplitudes calculated from the weighted mean amplitudes of the ground-based observations (see Table~3 and eq.~1 in P01 for more details).

\begin{table*}
\centering
\caption{\ross: frequencies, periods, and amplitudes determined from the new, 20~s cadence mode TESS observations. For comparison, we also list the predicted amplitudes in the TESS band pass, and finally the weighted mean amplitudes from ground-based measurements.}
\label{tabl:ross}
\begin{tabular}{cccccc}
\hline
\hline
  & Frequency [$\mu$Hz] & Period [s] & Ampl. [mma] & TESS pred. ampl. 825~nm [mma] & Weighted mean ampl. [mma] \\
\hline
$f_1$ & 4691.908(10) & 213.1329 & 3.98(24) & 4.34 & 6.86(4) \\
$f_2$ & 4699.977(00) & 212.7670 & 2.39(15) & 2.71 & 4.28(4) \\
$f_3$ & 3646.315(16) & 274.2495 & 2.56(22) & 2.79 & 4.41(4) \\
$f_4$ & 3639.328(17) & 274.7760 & 2.18(22) & 2.01 & 3.17(4) \\
\hline
\end{tabular}
\end{table*}

As shown in Table~\ref{tabl:ross}, we obtained close agreement with the ultrashort cadence mode amplitudes and the predicted amplitudes calculated from the weighted means of ground-based measurements. As detailed in P01, we can estimate the expected amplitudes in the TESS band pass using the known effective temperature of the star; see eq.~1 in P01. Therefore, the more precisely we know this effective temperature, the more accurate our estimate of the TESS amplitudes will be.


\subsection{\ec}
The star \ec\ (TIC\,033986466, $\alpha_{2000}=23^{\mathrm h}51^{\mathrm m}22^{\mathrm s}$, $\delta_{2000}=-24^{\mathrm d}08^{\mathrm m}17^{\mathrm s}$) was discovered by \citet{1993MNRAS.263L..13S}, who found that it shows a complex frequency structure with closely spaced frequencies and harmonics.

The previous TESS observations revealed the stochastic nature of the long-period modes in accordance with the results of \citet{2017ApJS..232...23H}. We also detected possible rotationally split frequencies and presented the corresponding rotation periods in P01.


We detected 14 frequencies above the detection limit in the light curve of the star obtained in 20\,s cadence mode. For completeness, we summarise our findings together with the results of the previous observations presented in P01 in Table~\ref{tabl:ecfreq2}.

\begin{figure*}
\centering
\includegraphics[width=\textwidth]{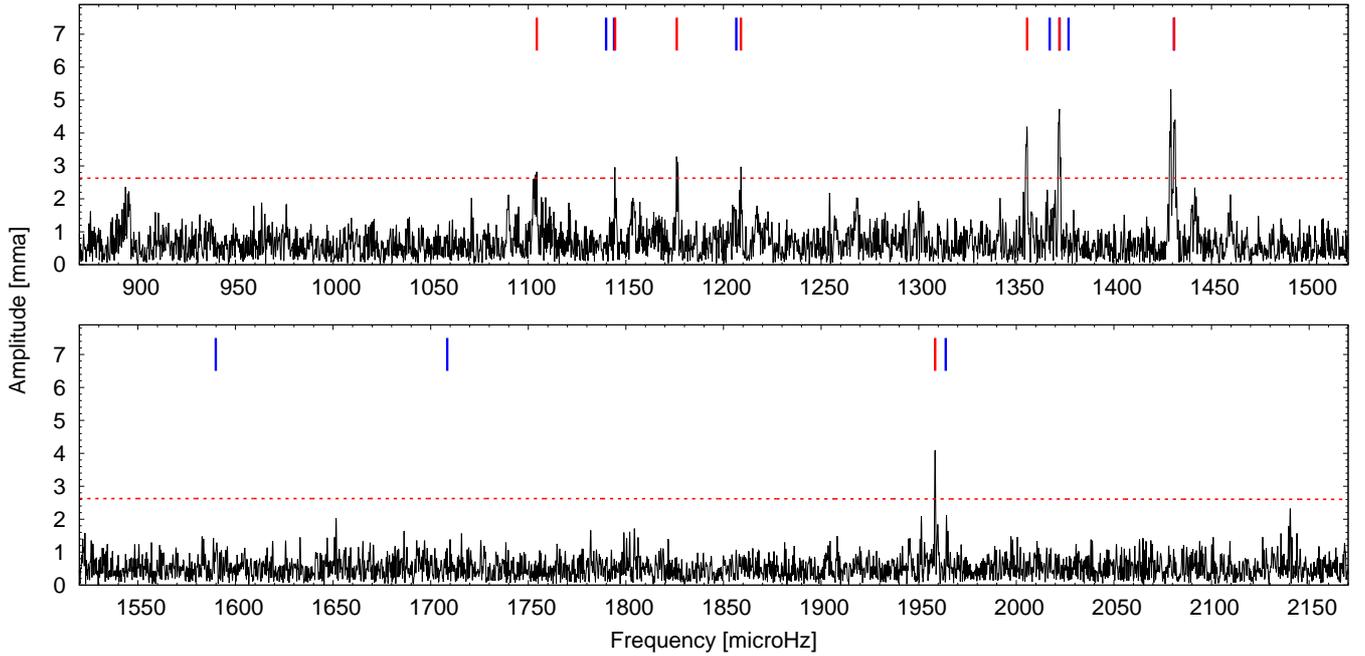}
\caption{\ec: Fourier transform of the ultrashort-cadence-mode light curve. Blue lines mark the frequencies detected in the previous TESS season (see P01), red lines denote the newly found frequencies, while the red dashed line corresponds to the $0.1\%$ FAP significance level.}
\label{fig:ecfreq}
\end{figure*}

\begin{table*}
\centering
\caption{\ec: results of the fit of the data set with 14 frequencies derived by the ultrashort cadence mode TESS observations (`This work').
The errors are formal uncertainties. P01 frequencies: the results of Lorentzian fits are marked with an asterisk. We also present a possible frequency identification in column 4. In the case of the supposed rotationally split frequencies, we mark the possible azimuthal order values. We use an `x' to mark the frequencies selected for asteroseismic period fits.}
\label{tabl:ecfreq2}
\begin{tabular}{llccllcc}
\hline
\hline
\multicolumn{3}{c}{This work} & & \multicolumn{3}{c}{P01 -- \citet{2020A&A...638A..82B}} & \\
\multicolumn{1}{c}{Frequency} & \multicolumn{1}{c}{Period} & \multicolumn{1}{c}{Ampl.} & & \multicolumn{1}{c}{Frequency} & \multicolumn{1}{c}{Period} & \multicolumn{1}{c}{Ampl.} & \multicolumn{1}{c}{Sel.}\\
\multicolumn{1}{c}{[$\mu$Hz]} & \multicolumn{1}{c}{[s]} & \multicolumn{1}{c}{[mma]} & & \multicolumn{1}{c}{[$\mu$Hz]} & \multicolumn{1}{c}{[s]} & \multicolumn{1}{c}{[mma]} & \\
\hline
1103.005(39) & 906.615 & 2.65(45) & & & & & \\
1104.293(37) & 905.557 & 2.84(45) & $f_1$ & & & & x \\
& & & $f_{2}^{-1/0?}$ & 1139.799(34) & 877.348 & 3.30(49) & x \\
1144.365(35) & 873.847 & 3.00(45) & $f_{2}^{0/+1?}$ & 1144.67(35)$^*$ & 873.614 & 1.61(30) & x \\
1175.997(31) & 850.342 & 3.31(45) & $f_3$ & & & & x \\
1208.947(35) & 827.166 & 2.99(45) & $f_{4}$ & 1206.495(39)$^*$ &    828.847 & 2.41(05) & x \\
1355.178(30) & 737.910 & 3.52(46) & & & & \\
1355.546(26) & 737.710 & 4.06(46) & $f_5$ & & & & x \\
& & & $f_{6}^{-1}$ & 1367.082(17) &   731.485 & 6.65(49) & \\
1371.708(24) & 729.018 & 4.30(46) & & & & \\
1372.103(22) & 728.808 & 4.67(46) & $f_{6}$ & 1372.116(35) &    728.802 & 3.15(49) & x \\
& & & $f_{6}^{+1}$ & 1376.720(39) &   726.364 & 2.81(49) & \\
1428.764(30) & 699.906 & 3.51(47) & & & & \\
1429.206(20) & 699.689 & 5.32(47) & & & & \\
1430.753(33) & 698.933 & 3.15(46) & $f_{7}$ & 1430.780(17) &        698.920 & 6.41(49) & x \\
1431.165(24) & 698.731 & 4.30(47) & & & & \\
& & & $f_{8}$ & 1589.850(37) &        628.990 & 2.96(49) & x \\
& & & $f_{9}$ & 1708.434(27) &        585.331 & 4.05(49) & x \\
1958.384(26) & 510.625 & 4.15(49) & $f_{10}^{-1/0?}$ & & & & x \\
 & & & $f_{10}^{0/+1?}$ & 1963.783(23) & 509.221 & 4.88(49) & x \\
\hline
\end{tabular}
\end{table*}

As shown in Table~\ref{tabl:ecfreq2}, the ultrashort-cadence data set reveals new pulsation-frequency components. We also note the presence of closely spaced frequencies, which might be the result of short-term amplitude, phase, or frequency variations above $\sim 700\,$s. This is  consistent with the findings of \citet{2017ApJS..232...23H}, who predicted stochastic behaviour of the periods above $\sim 800\,$s. Comparing the frequency contents of the \ec\ data sets obtained at different times, it is indeed clear that amplitude variations occur between the epochs.


\subsection{\bpm}
The star \bpm\ (TIC\,101014997, $\alpha_{2000}=03^{\mathrm h}43^{\mathrm m}29^{\mathrm s}$, $\delta_{2000}=-45^{\mathrm d}49^{\mathrm m}04^{\mathrm s}$) is known as a pulsator with a few independent modes, but shows several additional peaks: combination frequencies, near-subharmonic peaks, and frequencies emerging as a result of rotational frequency splitting. Amplitude variations from season to season are also known at some frequencies.

\begin{figure*}
\centering
\includegraphics[width=\textwidth]{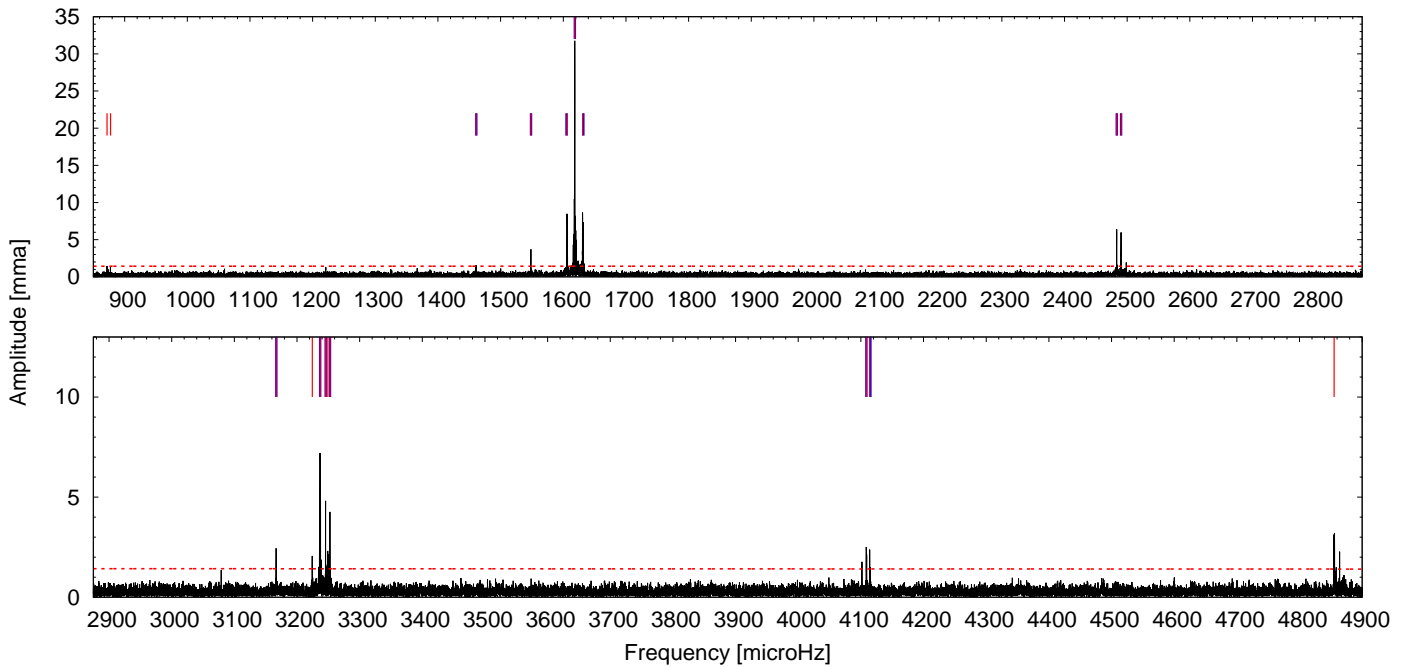}
\caption{\bpm: Fourier transform of the ultrashort-cadence TESS observations. For further explanation, we refer to the caption of Fig.~\ref{fig:ecfreq}}
\label{fig:bpmfreq}
\end{figure*}

Combining data from sectors 30 and 31, we find 43 significant peaks. However, these frequencies are not all independent pulsation modes; several of them form frequency groups, indicating short-term amplitude, phase, or frequency variations, or possible instrumental effects. In Table~\ref{tabl:bpmfreq2}, we summarise our findings, and also list the frequencies presented in P01. Figure~\ref{fig:bpmfreq} shows the Fourier transform of the ultrashort-cadence light curve of \bpm, in which we marked the new frequencies as well as the location of peaks identified in P01.

\begin{table*}
\centering
\caption{\bpm: Results of the fit of the data set derived from the ultrashort cadence mode TESS observations (`This work'). We use plus symbols to mark where closely spaced peaks around a larger-amplitude one exist. These may not represent additional independent pulsational frequencies, and therefore we do not include them in this list. The errors are formal uncertainties. We also present a possible frequency identification in the fourth column. We use an `x' to mark the frequencies selected for asteroseismic period fits.}
\label{tabl:bpmfreq2}
\begin{tabular}{lrrcrrrc}
\hline
\hline
\multicolumn{3}{c}{This work} & & \multicolumn{3}{c}{P01 -- \citet{2020A&A...638A..82B}} & \\
\multicolumn{1}{c}{Frequency [$\mu$Hz]} & \multicolumn{1}{c}{Period [s]} & \multicolumn{1}{c}{Amplitude [mma]} & & \multicolumn{1}{c}{Frequency [$\mu$Hz]} & \multicolumn{1}{c}{Period [s]} & \multicolumn{1}{c}{Amplitude [mma]} 
 & Sel. \\
\hline
 & & & $\Delta$ & 13.409(12) &        74577 &         3.08(32) & \\
871.799(18) & 1147.053 & 1.45(24) & $0.54f_4$ & & & & \\
877.261(18) & 1139.912 & 1.46(24) & $f_1$ & & & & x \\
1460.726(17) & 684.591 & 1.58(26) & $f_2$ & 1461.117(13) &  684.408 &       2.87(32) & x \\
1548.328(7) & 645.858 & 3.70(26) & $f_3$ & 1548.455(4) &   645.805 &       8.47(32) & x \\
1605.950(3)$^+$ & 622.684 & 8.68(27) & $f_4-\Delta$ & 1604.982(4) &      623.060 &       8.94(32) & \\
1618.374(1)$^+$ & 617.904 & 32.44(27) & $f_4$ & 1618.401(1) &   617.894 &       39.37(32) & x \\
1630.778(3)$^+$ & 613.204 & 8.50(27) & $f_4+\Delta$ & 1631.844(4) &      612.804 &       10.80(32) & \\
2483.254(4) & 402.697 & 6.40(30) & $f_x$ & 2483.203(7) & 402.706 & 5.06(32) & \\
2490.171(5)$^+$ & 401.579 & 6.09(30) & $1.54f_4$ & 2490.147(7) &       401.582 &       5.35(32) & \\
3166.716(12) & 315.785 & 2.40(32) & $f_3 + f_4$ & 3166.877(9) &     315.769 &       4.05(32) & \\
3224.310(13) & 310.144 & 2.15(32) & $2f_4-\Delta$ & & & & \\
3236.751(4)$^+$ & 308.952 & 7.21(33) & $2f_4$ & 3236.794(6) &  308.948 &       6.26(32) & \\
3245.653(6) & 308.104 & 4.85(32) & $f_y$ & 3245.648(11) & 308.105 & 3.41(32) & \\
3249.136(12)$^+$ & 307.774 & 2.26(32)  & $2f_4+\Delta$ & & & & \\
3252.603(7) & 307.446 & 4.24(32) & $f_w$ & 3252.590(12) & 307.447 & 3.16(32) & \\
4108.546(12)$^+$ & 243.395 & 2.51(35) & $2.54f_4$ & 4108.584(15) &      243.393 &       2.49(32) & \\
4114.021(12) & 243.071 & 2.34(35) & $f_z$ & 4115.083(16) & 243.009 & 2.41(32) & \\
4855.098(9)$^+$ & 205.969 & 2.55(37)  & $3f_4$ & & & & \\
5726.912(18) & 174.614 & 1.74(39)  & $3.54f_4$ & & & & \\
\hline
\end{tabular}
\end{table*}

The dominant peak is at $1618.4\,\mu$Hz, in agreement with the results of the previous observations. 
This peak and its first harmonic are also central components of two triplets with $\sim 13\,\mu$Hz frequency separation (denoted by $\Delta$ in Table~\ref{tabl:bpmfreq2}). Such triplets can be interpreted as rotational frequency splitting of an $\ell = 1$ mode.

There are several surprising frequencies denoted as $f_x, f_y, f_w$, and $f_z$. The origin of these peak is not clear, but it is worth mentioning that the spacings between $1.54f_4$ and $f_x$ ($\delta f = 6.92\,\mu$Hz), and $f_y$ and $f_w$ ($\delta f = 6.95\,\mu$Hz), and $2.54f_4$ and $f_z$ ($\delta f = 5.48\,\mu$Hz) are similar. We cannot rule out that there are additional independent frequencies among these peaks. Further observations may reveal their true nature.   

We note that there is another paper that deals with the frequency analyses of the TESS observations \citep{2023MNRAS.518.1448R}, and also reviews the outcomes of asterosesimological investigations. These authors came to a slightly different conclusion considering the mode identification of the observed frequencies, and used those periods for asteroseismology.

\subsection{\bpmb}
The star \bpmb\ (TIC\,102048288, $\alpha_{2000}=01^{\mathrm h}06^{\mathrm m}54^{\mathrm s}$, $\delta_{2000}=-46^{\mathrm d}08^{\mathrm m}54^{\mathrm s}$) shows complex pulsations in the ground-based observations. We analysed the ultrashort cadence data on \bpmb\ obtained in sector 29. We identified only five significant frequencies, of which two pairs form closely spaced doublets, suggesting that only three modes present in the new observations of \bpmb. Table~\ref{tabl:bpmbfreq2} summarises the frequencies found from the new TESS observations and also the frequencies published in P01.

\begin{table*}
\centering
\caption{\bpmb: Significant frequencies found in the sector 29 observations (`This work'). In the case of the P01 frequencies, they are results of Lorentzian fits to the frequency groups observed in the actual \bpmb\ data set. The central frequencies are marked with asterisks. We use an `x' to mark the frequencies selected for asteroseismic period fits.}
\label{tabl:bpmbfreq2}
\begin{tabular}{lrrlrrrc}
\hline
\hline
\multicolumn{3}{c}{This work} & & \multicolumn{3}{c}{P01 -- \citet{2020A&A...638A..82B}} & \\
\multicolumn{1}{c}{Frequency [$\mu$Hz]} & \multicolumn{1}{c}{Period [s]} & \multicolumn{1}{c}{Amplitude [mma]} & & \multicolumn{1}{c}{Frequency [$\mu$Hz]} & \multicolumn{1}{c}{Period [s]} & \multicolumn{1}{c}{Amplitude [mma]} & Sel. \\
\hline
 & & & $f_1$ & 1144.760(150)$^*$ & 873.546 & 1.39(03) & x \\
 & & & & 1204.487(021)$^*$ & 830.229 & 3.14(02) & \\
1208.880(33) & 827.212 & 3.37(48) & & & & \\
1209.997(27) & 826.448 & 4.17(48) & $f_2$ & & & & x \\
 & & & $f_3$ & 1264.164(093)$^*$ & 789.163 & 1.83(02) & x \\
1413.070(30) & 707.679 & 3.70(49) & $f_4$ & & & & x \\
1487.056(32) & 672.470 & 3.48(49) & $f_5$ & & & & x \\
1490.973(40) & 670.703 & 2.76(49) & & & & & \\
\hline
\end{tabular}
\end{table*}

\begin{figure*}
\centering
\includegraphics[width=\textwidth]{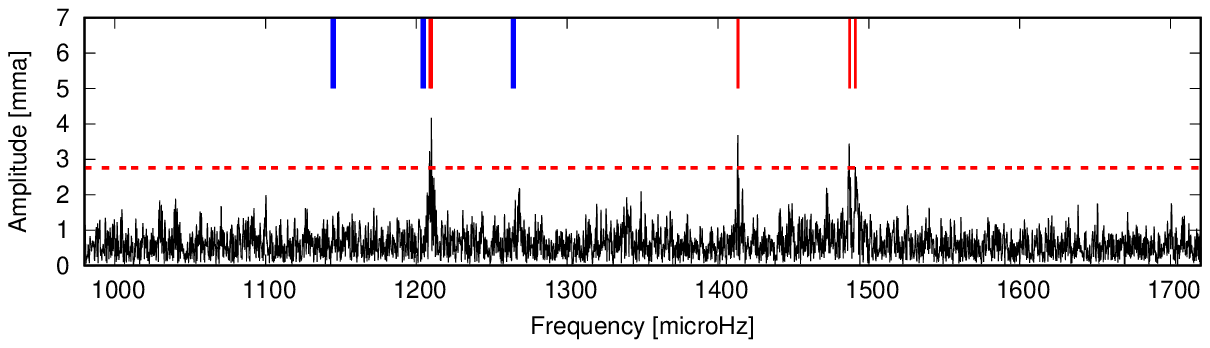}
\caption{\bpmb: Fourier transform of the ultrashort cadence TESS observations. For further explanation, we refer to the caption of Fig.~\ref{fig:ecfreq}.}
\label{fig:bpmbFT}
\end{figure*}

This star is an excellent example of the existence of large-amplitude variations from one observing season to another: new frequencies emerge above the significance level, while the amplitudes of formerly known frequencies decrease, or some of the frequencies cannot be detected at all, as shown in Fig.~\ref{fig:bpmbFT}.   


\subsection{\mct}
The star \mct\ (TIC~164772507, $\alpha_{2000}=01^{\mathrm h}47^{\mathrm m}22^{\mathrm s}$, $\delta_{2000}=-21^{\mathrm d}56^{\mathrm m}51^{\mathrm s}$) is another example of temporal changes in the pulsation properties of cool DAVs.


The standard pre-whitening process yielded 24 significant frequencies. However, similarly to \bpmb, there are frequency groups of closely spaced components in the Fourier transform, suggesting that amplitude, frequency, or phase variations occurred during the observations. We observe such behaviour in frequencies above 700\,s. We summarise our frequency identification together with previous results from P01 in Table~\ref{tabl:mctfreq}. We note that these are the results of an automated pre-whitening process, and there can be real pulsation modes, products of rotational frequency splittings, and peaks emerging from the stochastic nature of the long-period modes.

\begin{table*}
\centering
\caption{\mct: frequencies detected above the detection limit during the standard pre-whitening process (`This work'). P01: A complete list of frequencies derived from the \mct\ data set. We use an `x' to denote the frequencies selected for asteroseismic period fits.}
\label{tabl:mctfreq}
\begin{tabular}{lrrlrrrc}
\hline
\hline
\multicolumn{3}{c}{This work} & & \multicolumn{3}{c}{P01 -- \citet{2020A&A...638A..82B}} & \\
\multicolumn{1}{c}{Frequency [$\mu$Hz]} & \multicolumn{1}{c}{Period [s]} & \multicolumn{1}{c}{Amplitude [mma]} & & \multicolumn{1}{c}{Frequency [$\mu$Hz]} & \multicolumn{1}{c}{Period [s]} & \multicolumn{1}{c}{Amplitude [mma]} & Sel. \\
\hline
& & & $f_1$ & 955.854(044) & 1046.185 & 3.68(53) & x \\
1240.210(34) & 806.315 & 2.84(45) & & & & & \\
1241.333(29) & 805.585 & 3.38(45) & $f_2$ & & & & x \\
1243.911(35) & 803.916 & 2.76(45) & & & & & \\
& & & & 1253.099(045) & 798.021 & 3.63(1.89) & \\
& & & & 1257.085(063) & 795.491 & 5.29(53) & \\
& & & $f_3$ & 1257.462(052) & 795.252 & 6.40(88) & x \\
& & & $f_4?$ & 1297.472(039) & 770.730 & 4.13(62) & \\
& & & & 1305.221(045) & 766.154 & 3.72(55) & \\
& & & $f_5$ & 1306.979(033) & 765.123 & 5.21(89) & x \\
& & & & 1308.725(053) & 764.103 & 3.18(55) & \\
1313.363(23) & 761.404 & 4.20(45) & & & & & \\
1313.666(23) & 761.228 & 4.30(46) & $f_6?$ & & & & \\
1313.991(33) & 761.040 & 3.00(45) & & & & & \\
1314.336(37) & 760.840 & 2.67(45) & & & & & \\
& & & & 1317.983(046) & 758.735 & 3.52(1.90) & \\
1383.359(16) & 722.878 & 6.28(46) & & & & & \\
1384.254(37) & 722.411 & 2.63(46) & & & & & \\
1385.006(20) & 722.018 & 4.81(46) & & & & & \\
1385.522(12) & 721.750 & 8.48(46) & $f_7$ & & & & x \\
1386.095(37) & 721.451 & 2.62(46) & & & & & \\
1386.512(15) & 721.235 & 6.45(46) & & & & & \\
1388.434(35) & 720.236 & 2.78(46) & & & & & \\
1389.661(31) & 719.600 & 3.18(46) & & & & & \\
1390.189(23) & 719.327 & 4.21(46) & & & & & \\
1390.680(20) & 719.073 & 4.82(46) & & & & & \\
1391.311(23) & 718.746 & 4.22(46) & & & & & \\
1392.043(33) & 718.369 & 2.97(46) & & & & & \\
1394.614(28) & 717.044 & 3.45(46) & & & & & \\
1395.118(33) & 716.785 & 2.94(46) & & & & & \\
2200.133(35) & 454.518 & 2.83(51) & $f_8?$ & & & & \\
& & & & 2213.161(046) & 451.842 & 5.58(63) & \\
& & & & 2213.695(052) & 451.733 & 8.13(54) & \\
& & & $f_9$ & 2214.635(062) & 451.542 & 11.36(54) & x \\
& & & & 2214.985(055) & 451.470 & 8.35(53) & \\
& & & $f_{10}$ & 2407.200(015) & 415.420 & 10.90(53) & x \\
2697.888(00) & 370.660 & 2.49(42) & $f_{11}$ & & & & x \\ 
\hline
\end{tabular}
\end{table*}

\begin{figure*}
\centering
\includegraphics[width=\textwidth]{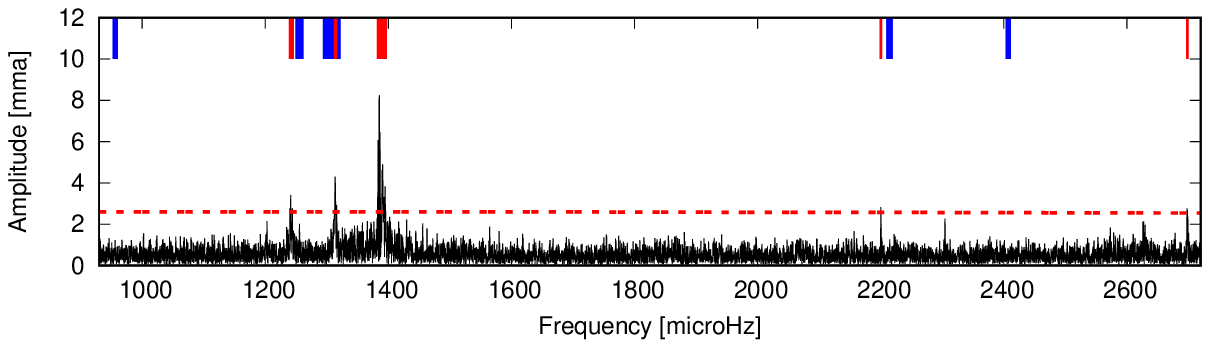}
\caption{\mct: Fourier transform of the ultrashort-cadence mode TESS observations. For further explanation, we refer to the caption of Fig.~\ref{fig:ecfreq}.}
\label{fig:mctFT}
\end{figure*}

It is conspicuous that there are no common frequencies between the two observing seasons, as if we observed two different stars. What we can find is groupings of periods around 805\,s, 795\,s, 765\,s, 761\,s, 720\,s, and 415\,s. In addition to these periods, there are a couple of peaks that do not belong to any group. These peaks are promising candidate eigenmodes for asteroseismic investigations. Figure~\ref{fig:mctFT} shows the Fourier transform of the  light curve from the 20\,s cadence mode, with the peaks found in the two different seasons marked.

\subsection{\lnt}
The star \lnt\ (TIC\,262872628, $\alpha_{2000}=14^{\mathrm h}33^{\mathrm m}08^{\mathrm s}$, $\delta_{2000}=-81^{\mathrm d}20^{\mathrm m}14^{\mathrm s}$) shows a simpler and much more stable pulsational behaviour than the previous two objects (\bpmb\ and \mct), with only a couple of eigenmodes showing rotational splittings. In the previous TESS observations presented in P01, we found only one frequency, the Nyquist alias of the 192.6\,s period. 

\begin{figure*}
\centering
\includegraphics[width=\textwidth]{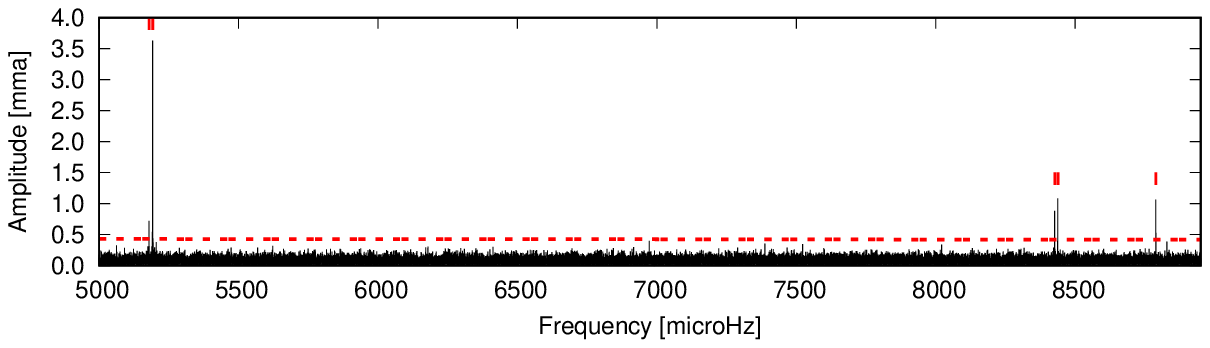}
\caption{\lnt: Fourier transform of the ultrashort-cadence mode TESS observations. Red lines denote the significant frequencies we detected using the sector 38--39 measurements.}
\label{fig:l19FT}
\end{figure*}

The analysis of the new data set for \lnt\ reveals the presence of five frequencies, listed in Table~\ref{tabl:lfreq}. For comparison, we also list the frequencies presented by \citet{2005ApJ...635.1239Y} in Table~\ref{tabl:lfreq}. Figure~\ref{fig:l19FT} shows the Fourier transform of the new TESS data.

\begin{table*}
\centering
\caption{\lnt: This work: Frequencies detected in the sector 38--39 data set. \citet{2005ApJ...635.1239Y} present frequencies and mode identification based on the results of a Whole Earth Telescope \citep{1990ApJ...361..309N} campaign. We use an `x' to mark the frequencies selected for asteroseismic period fits.}
\label{tabl:lfreq}
\begin{tabular}{lrrlrrrc}
\hline
\hline
\multicolumn{3}{c}{This work} & & \multicolumn{3}{c}{\citet{2005ApJ...635.1239Y}} & \\
\multicolumn{1}{c}{Frequency [$\mu$Hz]} & \multicolumn{1}{c}{Period [s]} & \multicolumn{1}{c}{Amplitude [mma]} & & \multicolumn{1}{c}{Frequency [$\mu$Hz]} & \multicolumn{1}{c}{Period [s]} & \multicolumn{1}{c}{Amplitude [mma]} & Sel. \\
\hline
5178.857(12) & 193.093 & 0.688(72) & $f_1^{-1}$ & 5179.362 & 193.074 & 0.973 & \\
5191.851(2) & 192.610 & 3.627(72) & $f_1$ & 5191.795 & 192.612 & 5.535 & x \\
 & & & $f_1^{+1}$ & 5204.160 & 192.154 & 1.216 & \\
8789.057(8) & 113.778 & 1.066(72) & $f_2$ & 8789.054 & 113.778 & 1.766 & x \\
 & & & $f_2^{+2}$ & 8828.619 & 113.268 & 0.271 & \\
8426.318(9) & 118.676 & 0.892(72) & $f_3^{-1}$ & 8426.324 & 118.676 & 1.191 & \\
8437.439(8) & 118.519 & 1.089(72) & $f_3$ & 8437.433 & 118.520 & 1.641 & x \\
 & & & $f_3^{+1}$ & 8448.580 & 118.363 & 0.339 & \\
 & & & $f_4$ & 2855.785 & 350.166 & 0.918 & \\
 & & & $f_4^{+1}$ & 2868.023 & 348.672 & 0.347 & \\
 & & & $f_5^{-1?}$ & 6954.415 & 143.794 & 0.228 & \\
 & & & $f_5$ & 6972.527 & 143.420 & 0.354 & \\
 & & & $f_5^{+1}$ & 6991.176 & 143.038 & 0.341 & \\
\hline
\end{tabular}
\end{table*}

\subsection{\hs}
The star \hs\ (TIC 277747736 $\alpha_{2000}=10^{\mathrm h}15^{\mathrm m}48^{\mathrm s}$, $\delta_{2000}=+03^{\mathrm d}06^{\mathrm m}48^{\mathrm s}$) is a DAV star observed in P01, but those data do not show light variations (NOV in P01). Previous ground-based observations showed that this star displays stable, short-period light variations similarly to \lnt.

\begin{figure*}
\centering
\includegraphics[width=\textwidth]{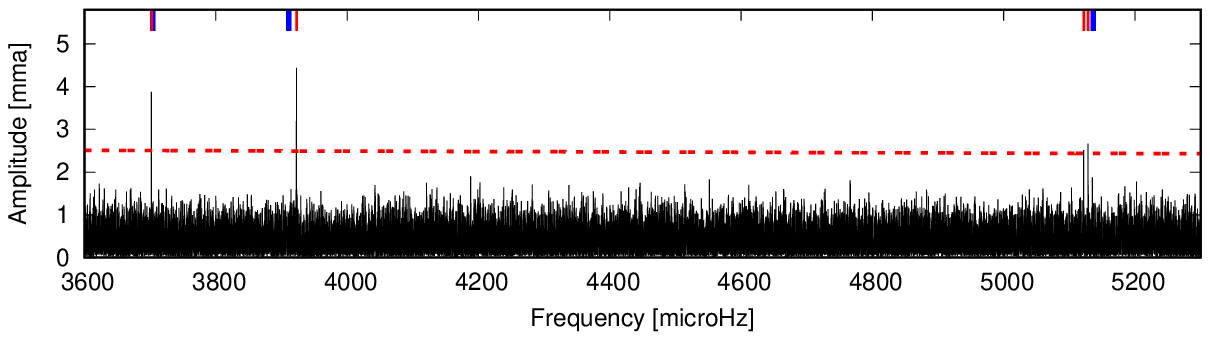}
\caption{\hs: Fourier transform of the ultrashort-cadence mode TESS observations. Red marks show the significant frequencies identified in this data, while blue marks show the location of pulsation components measured by \citet{2004ApJ...607..982M}.}
\label{fig:hsfreq}
\end{figure*}

TESS observed the star in three sectors in the ultrashort cadence mode. Using the Fourier analysis of the data set, we detect all three frequencies reported in the literature. Furthermore, we find a doublet in the Fourier transform at one of the frequencies, which may be the result of rotational frequency splitting. Table~\ref{tabl:hsfreq} summarises our findings together with the frequencies presented by \citet{2004ApJ...607..982M}, while Fig.~\ref{fig:hsfreq} shows the Fourier transform of the three-sector TESS observation.

\begin{table*}
\centering
\caption{\hs: Frequencies detected in the TESS data set, and the periods found by \citet{2004ApJ...607..982M} for comparison. We use an `x' to mark the frequencies selected for asteroseismic period fits.}
\label{tabl:hsfreq}
\begin{tabular}{lrrlrrrc}
\hline
\hline
\multicolumn{3}{c}{This work} & & \multicolumn{3}{c}{\citet{2004ApJ...607..982M}} & \\
\multicolumn{1}{c}{Frequency [$\mu$Hz]} & \multicolumn{1}{c}{Period [s]} & \multicolumn{1}{c}{Amplitude [mma]} & & \multicolumn{1}{c}{Frequency [$\mu$Hz]} & \multicolumn{1}{c}{Period [s]} & \multicolumn{1}{c}{Amplitude [mma]} & Sel. \\
\hline
3701.931(2) & 270.129 & 3.88(39) & $f_1$ & 3703.7 & 270.0 & 8.4 & x \\
3922.823(1) & 254.918 & 4.52(39) & $f_2$ & 3910.8 & 255.7 & 7.3 & x \\
5121.767(2) & 195.245 & 2.53(39) & $f_3^{-1,0?}$ & & & & \\
5128.417(2) & 194.992 & 2.70(39) & $f_3^{0,-1?}$ & 5136.1 & 194.7 & 5.8 & x \\
\hline
\end{tabular}
\end{table*}

\subsection{\he}
The star \he\, (TIC 382303117, $\alpha_{2000}=05^{\mathrm h}33^{\mathrm m}07^{\mathrm s}$, $\delta_{2000}=-56^{\mathrm d}03^{\mathrm m}53^{\mathrm s}$) is a known ZZ~Ceti variable showing pulsation periods of between about 400 and 1400\,s (see the details and references in P01). Thanks to its location near the southern ecliptic pole, the former, 120 second measurements investigated in P01 covered 13 sectors of observations. However, we cannot detect any significant peaks either by analysing the individual sectors or by investigating the whole data set. What we find is a burst-like feature in the light curve obtained in the second half of the sector 1 observations. The duration of the presumed burst was about 24\,h with an amplitude of nearly 20 per cent. We analysed both the first half of the sector 1 data set and the second half containing the burst-like event, respectively, and found that relatively large-amplitude peaks emerged around 2530\,$\mu$Hz in the latter case, which could be connected to the presumed outburst. However, we cannot detect such a large-amplitude event in the light curves of the other sectors.

\begin{figure*}
\centering
\includegraphics[width=\textwidth]{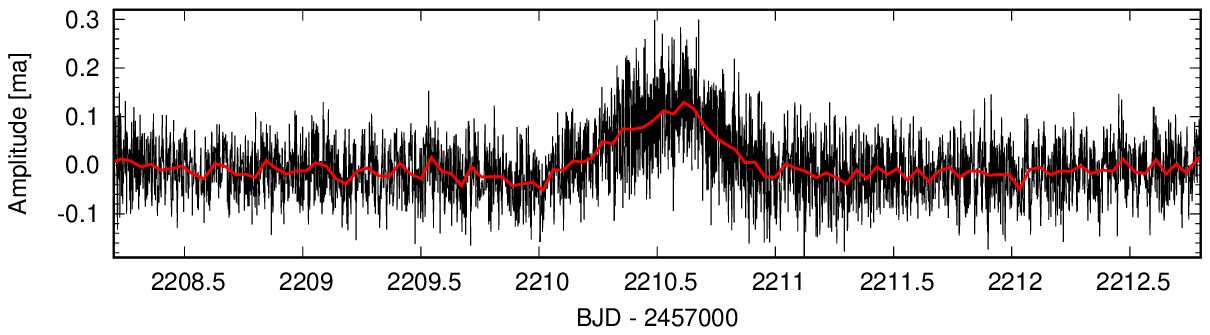}
\caption{\he: Portion of the light curve obtained by TESS during the 120 second cadence mode observations of  sector
33 (black line). The red line shows this light curve as binned, with a bin size of 30 points. The approximately 13 per cent flux emergence is clearly seen, and it takes about 1\,day.}
\label{fig:helc}
\end{figure*}

The star was observed in ten sectors in the ultrashort cadence mode. As in the former observations, we cannot detect any significant peaks in the Fourier transforms of the individual sectors or by the analysis of the whole data set. However, we investigated the 120 second cadence light curves as well, and find an outburst-like feature in the sector 33 data that is very similar to what was seen in the previous TESS observations. That is, the event repeated, which strongly suggests that \he\ is, in reality, a new outbursting cool DAV star. The flux increased by approximately 13 per cent during an interval of about 1\,day. The amplitude is smaller than the sudden 20 per cent increase detected in sector 1, but the one-day interval of the brightening phase is similar to the one observed previously. Figure~\ref{fig:helc} shows the corresponding part of the sector 33 light curve.

\begin{figure*}
\centering
\includegraphics[width=\textwidth]{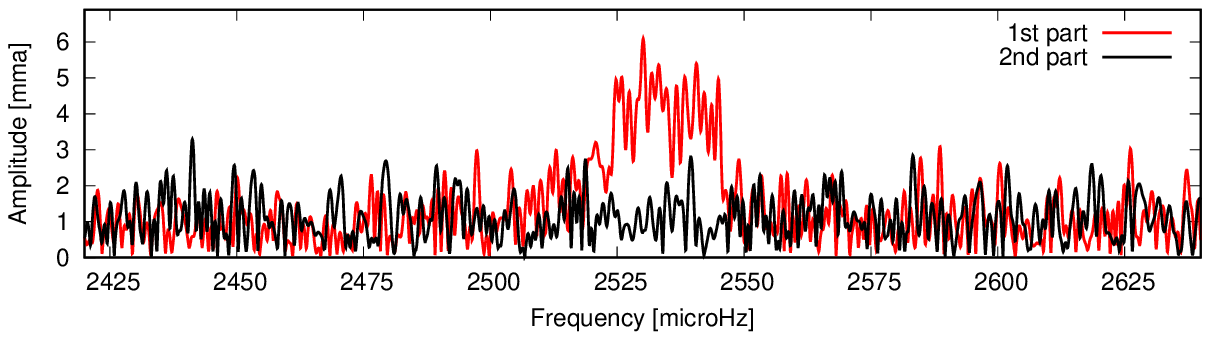}
\caption{\he: Red and black lines denote the Fourier transforms of the first and second halves of the light curve obtained in sector 33 and in 120 second cadence mode, respectively. The brightening effect in the light curve emerged during the first half of the observations.}
\label{fig:heparts}
\end{figure*}

As in P01, we checked the frequency content of the two halves of the sector 33 light curve. The brightening occurred during the first part of the observations, and as Fig.~\ref{fig:heparts} reveals, we see relatively large-amplitude peaks in the 2525--2545\,$\mu$Hz frequency domain in its Fourier transform. This phenomenon is in accordance with the former results. The peak of greatest amplitude can be found at 2530\,$\mu$Hz. All of this confirms that \he\ is a DAV star, which shows outburst events from time to time.

\begin{figure*}
\centering
\includegraphics[width=\textwidth]{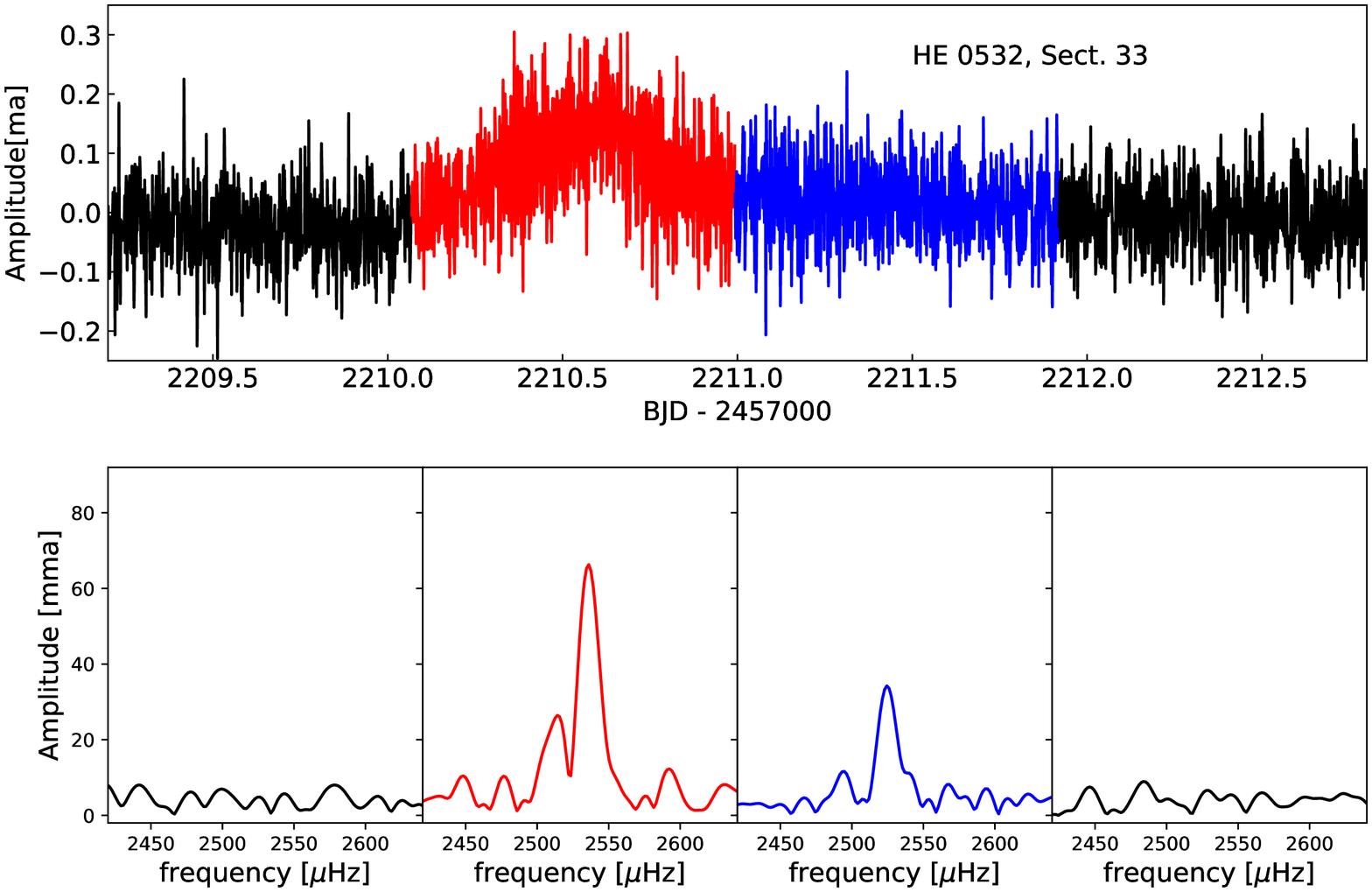}
\caption{\he: Top: 120\,s observations surrounding the outburst during sector 33. Colours correspond to 80\,000 second segments of the light curves for which the Fourier transforms are shown in the lower panels.}
\label{fig:hesect33pfts}
\end{figure*}

Looking more closely at the photometry immediately before, during, and after the outburst, we see in Fig.~\ref{fig:hesect33pfts} that there is no pulsation signal visible (above the noise level) in the 80\,000 second interval just prior to the outburst. During the main part of the outburst (seen in red), a significant periodicity at about 2535\,$\mu$Hz is apparent.  The next 80\,000 second segment, where the flaring is nearly gone, still shows some periodicity at 2525\,$\mu$Hz. The data are insufficient, given the short duration of the samples and the signal-to-noise ratio, to allow us to whether these are the same periodicity, or manifestations of amplitude or phase modulation or multiple periodicities. In either case, no evidence of variation remains after another 80\,000 seconds (the last segment in Fig.~\ref{fig:hesect33pfts}).

\begin{figure*}
\centering
\includegraphics[width=\textwidth]{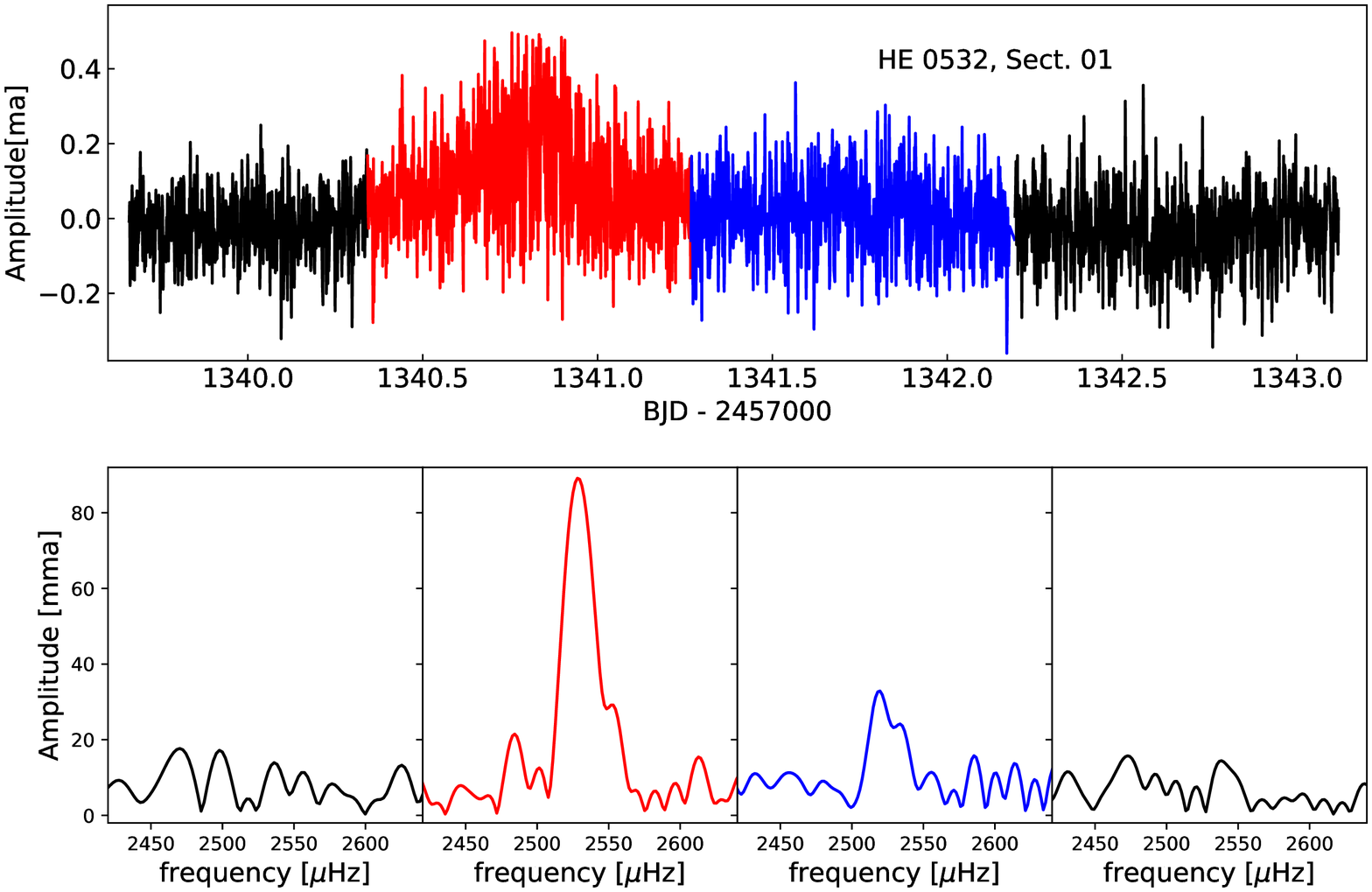}
\caption{\he: Same as Fig.~\ref{fig:hesect33pfts} but for data from sector 1.}
\label{fig:hesect01pfts}
\end{figure*}

\begin{figure*}
\centering
\includegraphics[width=\textwidth]{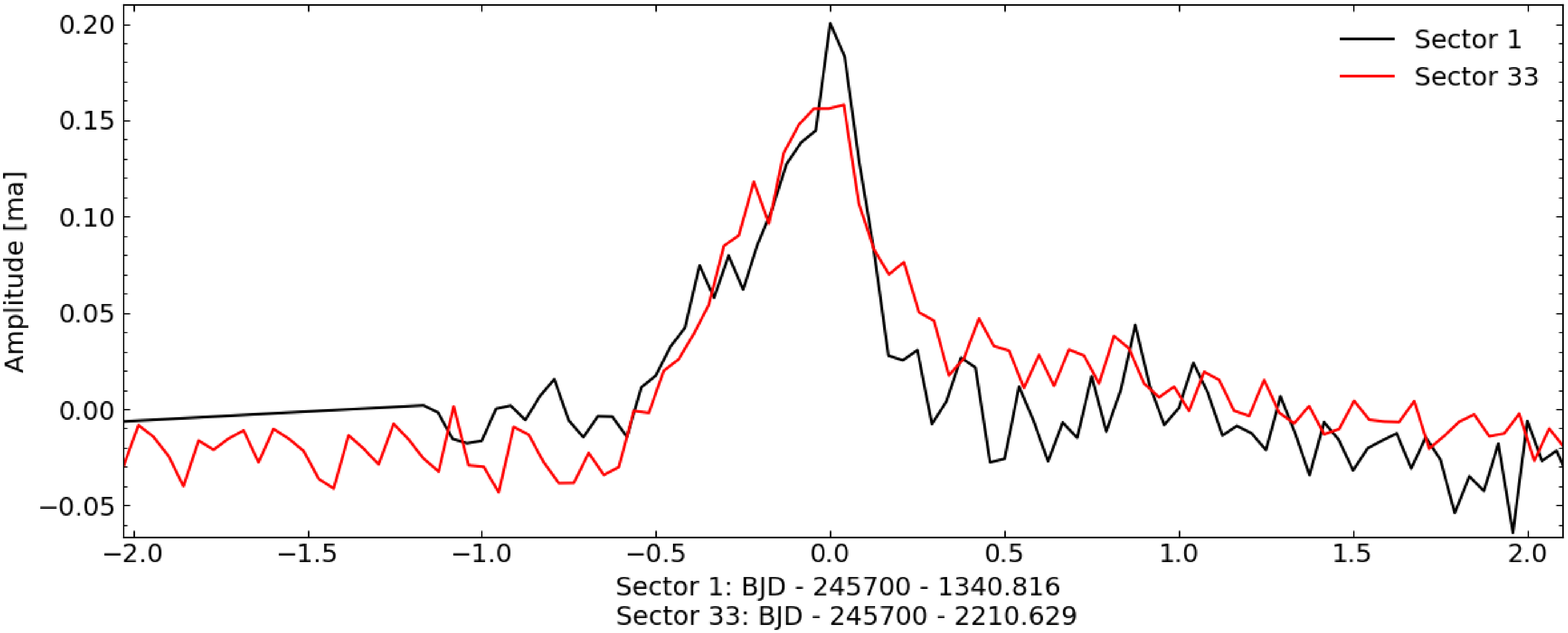}
\caption{\he: Bursts seen in \he\ in sector 1 and 33, aligned at the time of maximum amplitude. Here, 120\,s sampled data are smoothed using second-order polynomials over a range of 101 points. }
\label{fig:hesmoothedbursts}
\end{figure*}

We note very similar behaviour through the outburst in sector~1. Figure~\ref{fig:hesect01pfts} shows that the light curve through the outburst interval and the behaviour of the pulsation frequencies are nearly the same in the two sectors.  When the outbursts in sectors~1 and 33 are aligned in time (and the light curves smoothed; see Fig.~\ref{fig:hesmoothedbursts}), the bursts show similar rise times and durations. Given that only two bursts were seen in the extended monitoring of \he\  within the TESS extended coverage region, it appears that \he\  is an outbursting white dwarf with a very long interval between events.

\subsection{\wdj}

The star \wdj\ (TIC\,290653324, $\alpha_{2000}=09^{\mathrm h}25^{\mathrm m}12^{\mathrm s}$, $\delta_{2000}=+05^{\mathrm d}09^{\mathrm m}33^{\mathrm s}$) was classified as a NOV object in P01. We examined the new light curves, but do not find any periodic light variations at this time either.

\begin{figure*}
\centering
\includegraphics[width=\textwidth]{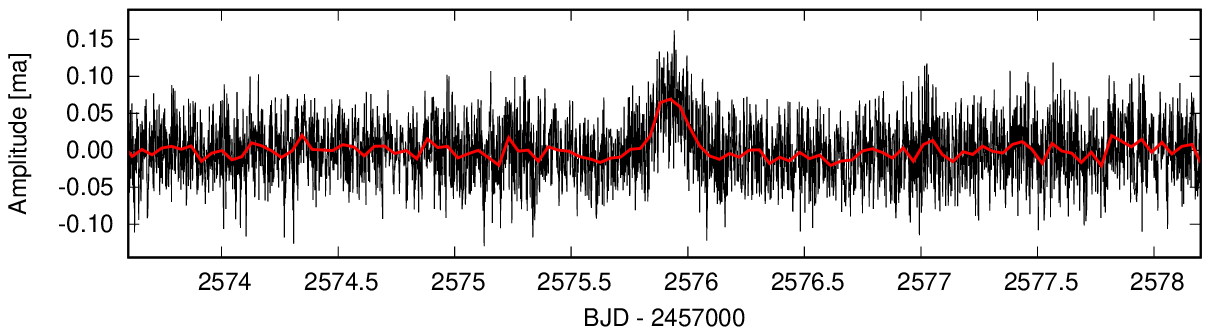}
\caption{\wdj: Portion of the light curve obtained by TESS during the sector 46 120-second cadence mode observations (black line). The red line shows this light curve as binned, with a bin size of 30 points. The maximum of the flux emergence is about 7 per cent, and the phenomenon takes about 0.5\,day.}
\label{fig:wdlc}
\end{figure*}

The star was observed in three sectors in ultrashort cadence mode. While we still do not find any sign of pulsational light variations, we identify an interesting brightening episode in the sector 46 light curve, similarly to the case of \he, and\  shown in Fig~\ref{fig:wdlc}. However, this is not intrinsic to the star; TESS coincidentally observed the minor planet (167578) 2004 BM$_{72}$ crossing the aperture of this target, producing a brightening episode of  several hours in duration, and rather similar in appearance to a weak outburst event. The Fourier spectrum of this light-curve segment also displays only noise.

\subsection{Discussion of the results of the frequency analyses}

Before turning to asteroseismology, it may be worth briefly discussing the results of the frequency analyses. The first question is how large is the difference between the detection threshold for the 120 second sampling mode described in P01 and the significance level obtained with the new 20 second sampling. The second refers to whether the different detection thresholds can explain the differences in the detected frequencies, or we are witnessing real amplitude changes considering the frequencies up to the Nyquist limit of the 120s cadence mode observations ($4167\,\mu$Hz). 

In Table~\ref{tabl:faps}, the detection thresholds described in P01 and those of this paper can be seen side by side. As Table~\ref{tabl:faps} shows, with the exception of two cases, we see values below $0.5$\,mma for the stars in terms of the differences in the detection limits. That is, with amplitude changes above the differences of the detection thresholds, we can witness real changes affecting the energy content of the individual pulsation modes. However, we must take care when interpreting peak amplitudes detected close to the Nyquist limit, as a smoothing or averaging effect occurs at these peaks that weakens the signals (see sect.~3.1.2 in P01).

\begin{table*}
\centering
\caption{Comparison of detection thresholds for the $0.1$ per cent FAP for 120s and 20s sampling modes.}
\label{tabl:faps}
\begin{tabular}{lcccc}
\hline
\hline
Object & TIC & {$0.1\%$\,FAP -- P01} & {$0.1\%$\,FAP -- this work} & $\Delta$\\
& & \multicolumn{1}{c}{(mma)} & \multicolumn{1}{c}{(mma)} & \multicolumn{1}{c}{(mma)}\\
\hline
\ross\ & 029854433 & 1.24 & 0.91 & 0.33\\
\ec\ & 033986466 & 2.83 & 2.60 & 0.23\\
\bpm\ & 101014997 & 1.89 & 1.47 & 0.42\\
\bpmb\ & 102048288 & 3.13 & 2.73 & 0.40\\
\mct\ & 164772507 & 3.00 & 2.57 & 0.43\\
\lnt\ & 262872628 & 0.65 & 0.44 & 0.21\\
\hs\ & 277747736 & 4.59 & 2.53 & 2.06\\
\wdj\ & 290653324 & 3.81 & 1.79 & 2.02\\
\he\ & 382303117 & 1.93 & 1.44 & 0.49\\
\hline
\end{tabular}
\end{table*}


\section{Preliminary asteroseismology}
\label{sect:seism}

Our asteroseismic investigations can be considered preliminary in the sense that (1) additional TESS observations are expected, which will complement the current period lists, (2) the step sizes used in the build of the model grids can be refined, and (3) the White Dwarf Evolution
Code (\textsc{wdec}) already has another version, with which the internal chemical profile characteristic of a given model can be parametrised in a different way to the one used here, and moreover, a model star can be built by changing up to 15 parameters. A more detailed asteroseismological analysis of the stars described in this paper is therefore possible.

For the modelling, we used the 2018 version of \textsc{wdec} \citep{2018AJ....155..187B}. This version uses the Modules for Experiments in Stellar Astrophysics (\textsc{mesa}; \citealt{2011ApJS..192....3P}, version r8118) opacity routines and equations of state for modelling.

Every evolutionary sequence starts with an $\sim100\,000$\,K polytrope, which is evolved down to a specified temperature. We search for a thermally relaxed solution to the stellar structure equations. Convection is treated within the framework of the mixing length theory \citep{1971A&A....12...21B}. The convective mixing length parameter  $\alpha$ was selected according to the prescription of \citet{2015ApJ...799..142T}. The $\ell=1$ and $2$ (nonradial) adiabatic pulsation modes for each model were calculated  \citep[see ][]{1989nos..book.....U}. We applied the \textsc{fitper} tool of \citet{2007PhDT........13K} to compare the periods of the observed ($P_i^{\mathrm{obs}}$) and the calculated ($P_i^{\mathrm{calc}}$) eigenmodes. The program calculates the root mean square ($\sigma_\mathrm{{rms}}$) value for every model as follows:

\begin{equation}
\sigma_\mathrm{{rms}} = \sqrt{\frac{\sum_{i=1}^{N} (P_i^{\mathrm{calc}} - P_i^{\mathrm{obs}})^2}{N}},
\label{equ1}
\end{equation}

\noindent where \textit{N} is the number of observed periods, and the goodness of the fits is characterised by this $\sigma_\mathrm{{rms}}$.

Knowing the luminosity of the models and the apparent visual magnitudes ($m_{\mathrm V}$) of the stars, given in the fourth US Naval Observatory CCD Astrograph Catalog \citep{2012yCat.1322....0Z}, we can calculate the seismic distances, as demonstrated by \citet{2019A&A...632A..42B} and \citet{2021A&A...651A..14B} for example.
We searched for a model solution amongst the best-fit models that
has a seismic distance close to (within $3\sigma$) the geometric distance provided by the Gaia space telescope's early third data release (EDR3; \citealt{2021A&A...649A...1G}). 

The model grid we used for the asteroseismic fits is the one we also used for the seismic investigations of the ZZ~Ceti star HS~1625+1231 \citep{2021AcA....71..281K}. Table~\ref{tabl:refined} lists the parameter space covered by this grid, and the corresponding step sizes. We scanned the parameter space as follows: $T_{\mathrm{eff}}$ (effective temperature), $M_*$ (stellar mass), $M_\mathrm{{env}}$ (the mass of the envelope, determined by the location of the base of the mixed helium and carbon layer), $M_\mathrm{H}$ (the mass of the hydrogen layer), $X_\mathrm{{He}}$ (the helium abundance in the C/He/H region), and $X_\mathrm{O}$ (the central oxygen abundance). We fixed the mass of the helium layer ($M_{\mathrm{He}}$) at $10^{-2}\,M_*$, which is the theoretical maximum for $M_{\mathrm{He}}$.

\begin{table}
\centering
\caption{Physical parameters varied while building the grid for asteroseismology. The step sizes applied are in parentheses.}
\label{tabl:refined}
\begin{tabular}{lr}
\hline
\hline
$T_{\mathrm{eff}}$ [K] & $10\,700 - 12\,300$ [100]\\
$M_*$ [$M_{\odot}$] & $0.56 - 0.74$ [0.01]\\
-log$(M_\mathrm{{env}}/M_*)$ & $1.5 - 1.9$ [0.1]\\
-log$(M_{\mathrm{He}}/M_*)$ & $2$ [fixed]\\
-log$(M_\mathrm{H}/M_*)$ & $4 - 9$ [$0.5$]\\ 
$X_\mathrm{{He}}$ & $0.5 - 0.9$ [0.1]\\
$X_\mathrm{O}$ & $0.5 - 0.9$ [0.1]\\
\hline
\end{tabular}
\end{table}

\begin{table*}
\centering
\caption{Effective temperature, surface gravity, mass, and apparent visual magnitude values of the stars showing pulsational light variations. The superscripts $G$ and $T$ denote the original sources of the effective temperature and surface gravity values. In the case of the errors determined by \citet{2011ApJ...743..138G}, these are around $200$\,K and $0.04 - 0.05$\,dex for the $T_{\mathrm{eff}}$ and log\,$g$ values, respectively, which means about $0.03\,M_{\odot}$ errors for the stellar masses. In the only case when the source of the atmospheric parameters was \citet{2011ApJ...730..128T}, the corresponding errors are $40$\,K and $0.01$\,dex.}
\label{tabl:seismo1}
\begin{tabular}{lrrrl}
\hline
\hline
\multicolumn{1}{c}{Star} & \multicolumn{1}{c}{$T_{\mathrm{eff}}$ [K]} & \multicolumn{1}{c}{log\,$g$ [dex]} & \multicolumn{1}{c}{$M_*$ [$M_{\odot}$]} & \multicolumn{1}{c}{$m_{\mathrm V}$ [mag]} \\
\hline
\ross & $12\,300^G$ & $8.03$ & $0.63$ & $14.148\pm0.03$ \\
\ec   & $11\,560^G$ & $8.09$ & $0.66$ & $15.256\pm0.01$ \\
\bpm  & $11\,500^G$ & $8.05$ & $0.64$ & $14.913\pm0.11$ \\
\bpmb & $11\,240^G$ & $8.16$ & $0.70$ & $15.394$ \\
\mct  & $11\,850^G$ & $8.15$ & $0.70$ & $15.150\pm0.07$ \\
\lnt  & $12\,070^G$ & $8.13$ & $0.69$ & $13.351\pm0.03$ \\
\hs   & $11\,630^T$ & $8.12$ & $0.68$ & $15.585\pm0.10$ \\
\hline
\end{tabular}
\end{table*}

When two similar periods were detected in this work and in P01, we adopted the amplitude-weighted average for the asteroseismic fits. The frequencies we used for asteroseismic modelling are indicated in the tables of Sect.~\ref{sect:analyses}.

Table~\ref{tabl:seismo1} summarises the spectroscopic $T_{\mathrm{eff}}$, surface gravity (log\,$g$), and the corresponding $M_*$ values, and also lists the catalogue values of $m_{\mathrm V}$. 
We note that we corrected the $T_{\mathrm{eff}}$ and log\,$g$ values of \citet{2011ApJ...743..138G} and \citet{2011ApJ...730..128T} according to the findings of \citet{2013A&A...559A.104T} based on radiation-hydrodynamics three-dimensional simulations of convective DA stellar atmospheres, as these authors originally used one-dimensional model atmospheres. We calculated the mass of the stars using their log\,$g,$ and taking into account the log\,$g$--$M_*$ relations  based on the evolutionary sequences published by \citet{2020ApJ...901...93B} as presented on the `synthetic colors and evolutionary sequences of hydrogen- and helium-atmosphere white dwarfs' webpage\footnote{\url{http://www.astro.umontreal.ca/~bergeron/CoolingModels}}.

Table~\ref{tabl:seismo2} summarises the physical parameters of our `best' asteroseismic-fit models.  
The subsequent sections briefly discuss our results of the seismic fits for the different stars.

\begin{table*}
\centering
\caption{Physical parameters of the asteroseismic best-fit models for the different stars. We also list the corresponding seismic distances ($d$), together with the geometric Gaia distances ($d_{\mathrm{Gaia}}$).}
\label{tabl:seismo2}
\tiny
\begin{tabular}{rrrrrrrrlr}
\hline
\hline
\multicolumn{1}{c}{$T_{\mathrm{eff}}$ [K]} & \multicolumn{1}{c}{$M_*$ [$M_{\odot}$]} & \multicolumn{1}{c}{-log$(M_\mathrm{{env}}/M_*)$} & \multicolumn{1}{c}{-log$(M_\mathrm{{He}}/M_*)$} & \multicolumn{1}{c}{-log$(M_\mathrm{{H}}/M_*)$} & \multicolumn{1}{c}{$X_\mathrm{{He}}$} & \multicolumn{1}{c}{$X_\mathrm{{O}}$} & \multicolumn{1}{c}{$\sigma_\mathrm{{rms}}$ [s])} & \multicolumn{1}{c}{$d$ [pc]} & \multicolumn{1}{c}{$d_{\mathrm{Gaia}}$ [pc]} \\
\hline
\multicolumn{10}{l}{\ec:} \\
12\,300 & 0.68 & 1.7 & 2.0 & 5.0 & 0.7 & 0.9 & 2.95 & 48.96$\pm$0.23 & 49.830$^{+0.088}_{-0.079}$ \\
\multicolumn{10}{l}{\bpm:} \\
11\,500 & 0.57 & 1.6 & 2.0 & 4.5 & 0.5 & 0.9 & 0.84 & 44.27$\pm$2.24 & 44.202$^{+0.045}_{-0.049}$ \\
\multicolumn{10}{l}{\bpmb:} \\
12\,100 & 0.72 & 1.7 & 2.0 & 4.5 & 0.9 & 0.9 & 0.84 & 49.17 & 49.743$^{+0.070}_{-0.076}$ \\
\multicolumn{10}{l}{\mct:} \\
11\,800 & 0.69 & 1.6 & 2.0 & 5.0 & 0.7 & 0.9 & 1.85 & 44.07$\pm$1.42 & 46.157$^{+0.065}_{-0.069}$ \\
\multicolumn{10}{l}{\lnt:} \\
12\,200 & 0.65 & 1.9 & 2.0 & 6.0 & 0.5 & 0.5 & 0.24 & 20.76$\pm$0.29 & 20.871$^{+0.007}_{-0.006}$ \\
\multicolumn{10}{l}{\hs:} \\
12\,000 & 0.59 & 1.6 & 2.0 & 9.0 & 0.6 & 0.7 & 0.33 & 60.43$\pm$2.78 & 60.248$^{+0.158}_{-0.166}$ \\
\hline
\end{tabular}
\end{table*}

\subsection{\ross}

For \ross, we detected two doublets in the TESS data. We know from ground-based measurements \citep{2015ApJ...815...56G} that these are actually side components of frequency triplets; however, we cannot detect the central components with the TESS data. For this reason, and considering the low number of possible modes, we did not perform asteroseismic period fits for this star.

\subsection{\ec}

Fortunately, \ec\ shows several pulsation frequencies, as listed in Table~\ref{tabl:ecfreq2}. We selected four ten-period solutions for the period fits to investigate. This was necessary because of the ambiguity in azimuthal order ($m$) in two frequency pairs, $f_2$ and $f_{10}$, as we need the $m=0$ frequencies for the asteroseismic fits.
We fixed the following three modes as $\ell = 1$: the 877 or 873\,s mode, the central component of the triplet $f_6$ (728\,s), and the period at 510 or 509\,s. The physical parameters of our selected model are listed in Table~\ref{tabl:seismo2}. The $\ell = 1$ modes of this solution are the 509.2, 728.8, 877.3, and 905.6\,s modes, while the rest of the periods at 585.3, 629.0, 698.9, 737.7, 827.9, and 850.3\,s are $\ell = 2$. Comparing the effective temperature and stellar mass values of the selected model with those calculated from the spectroscopic observations, it can be seen that we selected a model with an approximately  700\,K higher temperature, but with a similar mass.

\subsection{\bpm}

We considered the case of fitting with four modes ($f_1$ to $f_4$), and we assumed that the central component of the $f_4$ triplet at 617\,s is an $\ell = 1$ mode. Table~\ref{tabl:seismo2} shows a model solution belonging to this four-period fit, in which all modes at 617.9, 645.8, 684.5, and 1139.9\,s are $\ell = 1$. We note that the effective temperature of this model is the same, but its mass is 0.07 solar masses lower than the value obtained from spectroscopy.

\subsection{\bpmb}

We performed our fits with five periods, without any constraints on the $\ell$ values of the modes. Our selected model has three $\ell = 1$ modes at 672.5, 707.7, and 873.5\,s, while the $\ell = 2$ modes are the 789.2 and 827.2\,s ones. The stellar mass of this model has a similar value to that obtained from spectroscopy, but we see a large difference in the effective temperatures: our selected model is more than 800\,K hotter than predicted by spectroscopy.

\subsection{\mct}

The selected model solution has effective temperature and stellar mass in the vicinity of the spectroscopic values. We obtained three $\ell = 1$ modes (415.4, 451.5, 805.6\,s), while the rest of the periods fitted are found to be $\ell = 2$ (370.7, 721.2, 765.1, 795.3, and 1046.2\,s). 

\subsection{\lnt}

The star \lnt\ is not rich in known pulsation modes, as Table~\ref{tabl:lfreq} demonstrates. We could fit three periods only, while we fixed the $\ell$ value of the 192.6\,s mode as $\ell = 1$ based on the mode identification of \citet{2005ApJ...635.1239Y}.
In the case of our selected model, the two modes detected at 113.8 and 118.5\,s are $\ell = 2$. Regarding the effective temperature and stellar mass values, there are differences of only 130\,K and 0.04 solar masses  compared to the spectroscopic values.

\subsection{\hs}

For \hs, we fitted with three periods detected in the TESS data set.
The selected model gives two $\ell = 1$ (195.0, 254.9\,s) and one $\ell = 2$ (270.1\,s) solutions for the modes. In this case, the difference between the model and the spectroscopic effective temperature is only 370\,K, but in terms of the stellar mass, we see a larger deviation, of 0.09 solar masses.


\section{Summary and conclusions}
\label{sect:disc}

We focused on the results of the ultrashort cadence (20\,s) mode TESS observations of nine white dwarf stars known as ZZ~Ceti variables. These stars were presented in P01, but in that case the shortest cadence mode available was only 120\,s. The new TESS observations allow us to determine the pulsation modes without the Nyquist ambiguities of the 120\,s measurements, and to compare the frequency contents between the different observational cycles. We note that we investigated the measurements on 18 previously known ZZ~Ceti stars in P01, but more than half of them did not show periodic light variations in the TESS observations. We explain this as mainly resulting from the faintness of these targets, in combination with the crowding of the large TESS pixels.

Considering the new observations presented in this paper, we find significant pulsation frequencies in seven stars. Additionally, we detect one-one brightening episodes in the light curves of two targets. In the case of the first, \he, we detect a similar brightening phase to that detected  earlier and described in P01, that is, the phenomenon is recurring, implying that \he\ is most probably a new outbursting ZZ~Ceti star. However, the observed brightening of the other star, \wdj, was extrinsic, and caused by a passing minor planet crossing the photometric aperture of the star in the TESS images.

Using the effective temperature and surface gravity values for \he\ provided in \citet{2016IBVS.6184....1B} (11\,510\,K and 8.42\,dex, respectively); and placing it on the $T_{\mathrm{eff}}$--log\,$g$ diagram presented for example by \citet{2017ApJS..232...23H} (their Fig.~3), we see that the star can be found near the red edge of the empirical ZZ~Ceti instability strip. In this respect, \he\ is similar to other ZZ~Cetis, showing outbursts. 

Below, we briefly summarise our results for the different stars showing pulsations based on the new ultrashort cadence TESS observations:

\ross: We can clearly detect the four highest-amplitude frequencies using the ultrashort cadence data, without the Nyquist alias ambiguities of the 120\,s cadence observations.

\ec: The dominant frequency is different from the one detected in the previous TESS data. Some of the peaks seem to be unstable in amplitude, frequency, or phase, producing additional peaks around them.

\bpm: We find almost all the frequencies detected previously, as well as several additional frequencies.

\bpmb: Only new frequencies are detected, although one of them very near to another detected in the first TESS data set in P01.

\mct: Only new frequencies are detected, none of the old ones appeared in the new TESS data, almost as if we observed a different star.

\lnt: One already known frequency, and four additional frequencies in the 20\,s cadence data.

\hs: The first TESS data set did not show pulsational light variations, while in the new TESS data, we detect all three frequencies reported earlier in the literature. We also detect a new peak, which appears to be the result of rotational frequency splitting.

The newly detected frequencies impose stronger constraints on asteroseismological modelling. We performed a preliminary asteroseismic analysis of the stars that show pulsational light variations, as the ultimate goal of our efforts to detect as many pulsation modes as we possibly can is to learn more about the internal structure of the target stars and their non-pulsating counterparts, and about the dynamical processes operating in them. We succeeded in finding models with parameters in the vicinity of the Gaia geometric distances.
Here, we demonstrate the high value of the new, ultrashort cadence mode observations in studying white dwarf variables, and the continuation of these measurements could be extremely valuable to the white dwarf community.  


\begin{acknowledgements}

The authors thank the anonymous referee for the constructive comments and recommendations on the manuscript.

The authors also thank S.~O. Kepler (Instituto de F\'{\i}sica, Universidade Federal do Rio Grande do Sul, Brazil) for sharing his results on the Fourier analyses of the light curves.

ZsB and \'AS acknowledge the financial support of the Lend\"ulet Program of the Hungarian Academy of Sciences, projects No. LP2018-7/2022, and LP2012-31. This research was supported by the KKP-137523 `SeismoLab' \'Elvonal grant of the Hungarian Research, Development and Innovation Office (NKFIH).

ZsB acknowledges the support by the J\'anos Bolyai Research Scholarship of the Hungarian Academy of Sciences.

This research was supported in part by the U.S. National Science Foundation under Grant No. NSF PHY-1748958, enabling SDK's participation in the KITP program on white dwarfs in 2022.

This paper includes data collected with the TESS mission, obtained from the MAST data archive at the Space Telescope Science Institute (STScI). Funding for the TESS mission is provided by the NASA Explorer Program. STScI is operated by the Association of Universities for Research in Astronomy, Inc., under NASA contract NAS 5–26555.

\end{acknowledgements}



\bibliographystyle{aa} 
\bibliography{9DAVs} 

\begin{thebibliography}{44}
\expandafter\ifx\csname natexlab\endcsname\relax\def\natexlab#1{#1}\fi

\bibitem[{{Althaus} {et~al.}(2010){Althaus}, {C{\'o}rsico}, {Isern}, \&
  {Garc{\'{\i}}a-Berro}}]{2010A&ARv..18..471A}
{Althaus}, L.~G., {C{\'o}rsico}, A.~H., {Isern}, J., \& {Garc{\'{\i}}a-Berro},
  E. 2010, \aapr, 18, 471

\bibitem[{{B{\'e}dard} {et~al.}(2020){B{\'e}dard}, {Bergeron}, {Brassard}, \&
  {Fontaine}}]{2020ApJ...901...93B}
{B{\'e}dard}, A., {Bergeron}, P., {Brassard}, P., \& {Fontaine}, G. 2020, \apj,
  901, 93

\bibitem[{{Bell} {et~al.}(2019){Bell}, {C{\'o}rsico}, {Bischoff-Kim},
  {Althaus}, {Bradley}, {Calcaferro}, {Montgomery}, {Uzundag}, {Baran},
  {Bogn{\'a}r}, {Charpinet}, {Ghasemi}, \& {Hermes}}]{2019A&A...632A..42B}
{Bell}, K.~J., {C{\'o}rsico}, A.~H., {Bischoff-Kim}, A., {et~al.} 2019, \aap,
  632, A42

\bibitem[{{Bell} {et~al.}(2015){Bell}, {Hermes}, {Bischoff-Kim}, {Moorhead},
  {Montgomery}, {{\O}stensen}, {Castanheira}, \&
  {Winget}}]{2015ApJ...809...14B}
{Bell}, K.~J., {Hermes}, J.~J., {Bischoff-Kim}, A., {et~al.} 2015, \apj, 809,
  14

\bibitem[{{Bell} {et~al.}(2016){Bell}, {Hermes}, {Montgomery}, {Gentile
  Fusillo}, {Raddi}, {G{\"a}nsicke}, {Winget}, {Dennihy}, {Gianninas},
  {Tremblay}, {Chote}, \& {Winget}}]{2016ApJ...829...82B}
{Bell}, K.~J., {Hermes}, J.~J., {Montgomery}, M.~H., {et~al.} 2016, \apj, 829,
  82

\bibitem[{{Bell} {et~al.}(2017){Bell}, {Hermes}, {Montgomery}, {Winget},
  {Gentile Fusillo}, {Raddi}, \& {G{\"a}nsicke}}]{2017ASPC..509..303B}
{Bell}, K.~J., {Hermes}, J.~J., {Montgomery}, M.~H., {et~al.} 2017, in
  Astronomical Society of the Pacific Conference Series, Vol. 509, 20th
  European White Dwarf Workshop, ed. P.-E. {Tremblay}, B.~{Gaensicke}, \&
  T.~{Marsh}, 303

\bibitem[{{Bischoff-Kim} \& {Montgomery}(2018)}]{2018AJ....155..187B}
{Bischoff-Kim}, A. \& {Montgomery}, M.~H. 2018, \aj, 155, 187

\bibitem[{{Bogn{\'a}r} {et~al.}(2021){Bogn{\'a}r}, {Kalup}, \&
  {S{\'o}dor}}]{2021A&A...651A..14B}
{Bogn{\'a}r}, Z., {Kalup}, C., \& {S{\'o}dor}, {\'A}. 2021, \aap, 651, A14

\bibitem[{{Bogn{\'a}r} {et~al.}(2020){Bogn{\'a}r}, {Kawaler}, {Bell},
  {Schrandt}, {Baran}, {Bradley}, {Hermes}, {Charpinet}, {Handler}, {Mullally},
  {Murphy}, {Raddi}, {S{\'o}dor}, {Tremblay}, {Uzundag}, \&
  {Zong}}]{2020A&A...638A..82B}
{Bogn{\'a}r}, Z., {Kawaler}, S.~D., {Bell}, K.~J., {et~al.} 2020, \aap, 638,
  A82

\bibitem[{{Bogn\'ar} \& {S\'odor}(2016)}]{2016IBVS.6184....1B}
{Bogn\'ar}, Z. \& {S\'odor}, A. 2016, Information Bulletin on Variable Stars,
  6184, 1

\bibitem[{{Bohm} \& {Cassinelli}(1971)}]{1971A&A....12...21B}
{Bohm}, K.~H. \& {Cassinelli}, J. 1971, \aap, 12, 21

\bibitem[{{Brickhill}(1991)}]{1991MNRAS.251..673B}
{Brickhill}, A.~J. 1991, \mnras, 251, 673

\bibitem[{{Charpinet} {et~al.}(2019){Charpinet}, {Brassard}, {Fontaine}, {Van
  Grootel}, {Zong}, {Giammichele}, {Heber}, {Bogn{\'a}r}, {Geier}, {Green},
  {Hermes}, {Kilkenny}, {{\O}stensen}, {Pelisoli}, {Silvotti}, {Telting},
  {Vu{\v{c}}kovi{\'c}}, {Worters}, {Baran}, {Bell}, {Bradley}, {Debes},
  {Kawaler}, {Ko{\l}aczek-Szyma{\'n}ski}, {Murphy}, {Pigulski}, {S{\'o}dor},
  {Uzundag}, {Handberg}, {Kjeldsen}, {Ricker}, \&
  {Vanderspek}}]{2019A&A...632A..90C}
{Charpinet}, S., {Brassard}, P., {Fontaine}, G., {et~al.} 2019, \aap, 632, A90

\bibitem[{{C{\'o}rsico}(2020)}]{2020FrASS...7...47C}
{C{\'o}rsico}, A.~H. 2020, Frontiers in Astronomy and Space Sciences, 7, 47

\bibitem[{{C{\'o}rsico} {et~al.}(2019){C{\'o}rsico}, {Althaus}, {Miller
  Bertolami}, \& {Kepler}}]{2019A&ARv..27....7C}
{C{\'o}rsico}, A.~H., {Althaus}, L.~G., {Miller Bertolami}, M.~M., \& {Kepler},
  S.~O. 2019, \aapr, 27, 7

\bibitem[{{Dolez} \& {Vauclair}(1981)}]{1981A&A...102..375D}
{Dolez}, N. \& {Vauclair}, G. 1981, \aap, 102, 375

\bibitem[{{Fontaine} \& {Brassard}(2008)}]{2008PASP..120.1043F}
{Fontaine}, G. \& {Brassard}, P. 2008, \pasp, 120, 1043

\bibitem[{{Gaia Collaboration} {et~al.}(2021){Gaia Collaboration}, {Brown},
  {Vallenari}, {Prusti}, {de Bruijne}, {Babusiaux}, {Biermann}, {Creevey},
  {Evans}, {Eyer}, {Hutton}, {Jansen}, {Jordi}, {Klioner}, {Lammers},
  {Lindegren}, {Luri}, {Mignard}, {Panem}, {Pourbaix}, {Randich}, {Sartoretti},
  {Soubiran}, {Walton}, {Arenou}, {Bailer-Jones}, {Bastian}, {Cropper},
  {Drimmel}, {Katz}, {Lattanzi}, {van Leeuwen}, {Bakker}, {Cacciari},
  {Casta{\~n}eda}, {De Angeli}, {Ducourant}, {Fabricius}, {Fouesneau},
  {Fr{\'e}mat}, {Guerra}, {Guerrier}, {Guiraud}, {Jean-Antoine Piccolo},
  {Masana}, {Messineo}, {Mowlavi}, {Nicolas}, {Nienartowicz}, {Pailler},
  {Panuzzo}, {Riclet}, {Roux}, {Seabroke}, {Sordo}, {Tanga}, {Th{\'e}venin},
  {Gracia-Abril}, {Portell}, {Teyssier}, {Altmann}, {Andrae}, {Bellas-Velidis},
  {Benson}, {Berthier}, {Blomme}, {Brugaletta}, {Burgess}, {Busso}, {Carry},
  {Cellino}, {Cheek}, {Clementini}, {Damerdji}, {Davidson}, {Delchambre},
  {Dell'Oro}, {Fern{\'a}ndez-Hern{\'a}ndez}, {Galluccio}, {Garc{\'\i}a-Lario},
  {Garcia-Reinaldos}, {Gonz{\'a}lez-N{\'u}{\~n}ez}, {Gosset}, {Haigron},
  {Halbwachs}, {Hambly}, {Harrison}, {Hatzidimitriou}, {Heiter},
  {Hern{\'a}ndez}, {Hestroffer}, {Hodgkin}, {Holl}, {Jan{\ss}en}, {Jevardat de
  Fombelle}, {Jordan}, {Krone-Martins}, {Lanzafame}, {L{\"o}ffler}, {Lorca},
  {Manteiga}, {Marchal}, {Marrese}, {Moitinho}, {Mora}, {Muinonen}, {Osborne},
  {Pancino}, {Pauwels}, {Petit}, {Recio-Blanco}, {Richards}, {Riello},
  {Rimoldini}, {Robin}, {Roegiers}, {Rybizki}, {Sarro}, {Siopis}, {Smith},
  {Sozzetti}, {Ulla}, {Utrilla}, {van Leeuwen}, {van Reeven}, {Abbas}, {Abreu
  Aramburu}, {Accart}, {Aerts}, {Aguado}, {Ajaj}, {Altavilla}, {{\'A}lvarez},
  {{\'A}lvarez Cid-Fuentes}, {Alves}, {Anderson}, {Anglada Varela}, {Antoja},
  {Audard}, {Baines}, {Baker}, {Balaguer-N{\'u}{\~n}ez}, {Balbinot}, {Balog},
  {Barache}, {Barbato}, {Barros}, {Barstow}, {Bartolom{\'e}}, {Bassilana},
  {Bauchet}, {Baudesson-Stella}, {Becciani}, {Bellazzini}, {Bernet}, {Bertone},
  {Bianchi}, {Blanco-Cuaresma}, {Boch}, {Bombrun}, {Bossini}, {Bouquillon},
  {Bragaglia}, {Bramante}, {Breedt}, {Bressan}, {Brouillet}, {Bucciarelli},
  {Burlacu}, {Busonero}, {Butkevich}, {Buzzi}, {Caffau}, {Cancelliere},
  {C{\'a}novas}, {Cantat-Gaudin}, {Carballo}, {Carlucci}, {Carnerero},
  {Carrasco}, {Casamiquela}, {Castellani}, {Castro-Ginard}, {Castro Sampol},
  {Chaoul}, {Charlot}, {Chemin}, {Chiavassa}, {Cioni}, {Comoretto}, {Cooper},
  {Cornez}, {Cowell}, {Crifo}, {Crosta}, {Crowley}, {Dafonte}, {Dapergolas},
  {David}, {David}, {de Laverny}, {De Luise}, {De March}, {De Ridder}, {de
  Souza}, {de Teodoro}, {de Torres}, {del Peloso}, {del Pozo}, {Delbo},
  {Delgado}, {Delgado}, {Delisle}, {Di Matteo}, {Diakite}, {Diener},
  {Distefano}, {Dolding}, {Eappachen}, {Edvardsson}, {Enke}, {Esquej}, {Fabre},
  {Fabrizio}, {Faigler}, {Fedorets}, {Fernique}, {Fienga}, {Figueras},
  {Fouron}, {Fragkoudi}, {Fraile}, {Franke}, {Gai}, {Garabato},
  {Garcia-Gutierrez}, {Garc{\'\i}a-Torres}, {Garofalo}, {Gavras}, {Gerlach},
  {Geyer}, {Giacobbe}, {Gilmore}, {Girona}, {Giuffrida}, {Gomel}, {Gomez},
  {Gonzalez-Santamaria}, {Gonz{\'a}lez-Vidal}, {Granvik},
  {Guti{\'e}rrez-S{\'a}nchez}, {Guy}, {Hauser}, {Haywood}, {Helmi}, {Hidalgo},
  {Hilger}, {H{\l}adczuk}, {Hobbs}, {Holland}, {Huckle}, {Jasniewicz},
  {Jonker}, {Juaristi Campillo}, {Julbe}, {Karbevska}, {Kervella}, {Khanna},
  {Kochoska}, {Kontizas}, {Kordopatis}, {Korn}, {Kostrzewa-Rutkowska},
  {Kruszy{\'n}ska}, {Lambert}, {Lanza}, {Lasne}, {Le Campion}, {Le Fustec},
  {Lebreton}, {Lebzelter}, {Leccia}, {Leclerc}, {Lecoeur-Taibi}, {Liao},
  {Licata}, {Lindstr{\o}m}, {Lister}, {Livanou}, {Lobel}, {Madrero Pardo},
  {Managau}, {Mann}, {Marchant}, {Marconi}, {Marcos Santos}, {Marinoni},
  {Marocco}, {Marshall}, {Martin Polo}, {Mart{\'\i}n-Fleitas}, {Masip},
  {Massari}, {Mastrobuono-Battisti}, {Mazeh}, {McMillan}, {Messina},
  {Michalik}, {Millar}, {Mints}, {Molina}, {Molinaro}, {Moln{\'a}r},
  {Montegriffo}, {Mor}, {Morbidelli}, {Morel}, {Morris}, {Mulone}, {Munoz},
  {Muraveva}, {Murphy}, {Musella}, {Noval}, {Ord{\'e}novic}, {Orr{\`u}},
  {Osinde}, {Pagani}, {Pagano}, {Palaversa}, {Palicio}, {Panahi}, {Pawlak},
  {Pe{\~n}alosa Esteller}, {Penttil{\"a}}, {Piersimoni}, {Pineau}, {Plachy},
  {Plum}, {Poggio}, {Poretti}, {Poujoulet}, {Pr{\v{s}}a}, {Pulone}, {Racero},
  {Ragaini}, {Rainer}, {Raiteri}, {Rambaux}, {Ramos}, {Ramos-Lerate}, {Re
  Fiorentin}, {Regibo}, {Reyl{\'e}}, {Ripepi}, {Riva}, {Rixon}, {Robichon},
  {Robin}, {Roelens}, {Rohrbasser}, {Romero-G{\'o}mez}, {Rowell}, {Royer},
  {Rybicki}, {Sadowski}, {Sagrist{\`a} Sell{\'e}s}, {Sahlmann}, {Salgado},
  {Salguero}, {Samaras}, {Sanchez Gimenez}, {Sanna}, {Santove{\~n}a},
  {Sarasso}, {Schultheis}, {Sciacca}, {Segol}, {Segovia}, {S{\'e}gransan},
  {Semeux}, {Shahaf}, {Siddiqui}, {Siebert}, {Siltala}, {Slezak}, {Smart},
  {Solano}, {Solitro}, {Souami}, {Souchay}, {Spagna}, {Spoto}, {Steele},
  {Steidelm{\"u}ller}, {Stephenson}, {S{\"u}veges}, {Szabados}, {Szegedi-Elek},
  {Taris}, {Tauran}, {Taylor}, {Teixeira}, {Thuillot}, {Tonello}, {Torra},
  {Torra}, {Turon}, {Unger}, {Vaillant}, {van Dillen}, {Vanel}, {Vecchiato},
  {Viala}, {Vicente}, {Voutsinas}, {Weiler}, {Wevers}, {Wyrzykowski}, {Yoldas},
  {Yvard}, {Zhao}, {Zorec}, {Zucker}, {Zurbach}, \&
  {Zwitter}}]{2021A&A...649A...1G}
{Gaia Collaboration}, {Brown}, A.~G.~A., {Vallenari}, A., {et~al.} 2021, \aap,
  649, A1

\bibitem[{{Giammichele} {et~al.}(2015){Giammichele}, {Fontaine}, {Bergeron},
  {Brassard}, {Charpinet}, {Pfeiffer}, \& {Vauclair}}]{2015ApJ...815...56G}
{Giammichele}, N., {Fontaine}, G., {Bergeron}, P., {et~al.} 2015, \apj, 815, 56

\bibitem[{{Gianninas} {et~al.}(2011){Gianninas}, {Bergeron}, \&
  {Ruiz}}]{2011ApJ...743..138G}
{Gianninas}, A., {Bergeron}, P., \& {Ruiz}, M.~T. 2011, \apj, 743, 138

\bibitem[{{Goldreich} \& {Wu}(1999)}]{1999ApJ...511..904G}
{Goldreich}, P. \& {Wu}, Y. 1999, \apj, 511, 904

\bibitem[{{Hermes} {et~al.}(2017){Hermes}, {G{\"a}nsicke}, {Kawaler}, {Greiss},
  {Tremblay}, {Gentile Fusillo}, {Raddi}, {Fanale}, {Bell}, {Dennihy}, {Fuchs},
  {Dunlap}, {Clemens}, {Montgomery}, {Winget}, {Chote}, {Marsh}, \&
  {Redfield}}]{2017ApJS..232...23H}
{Hermes}, J.~J., {G{\"a}nsicke}, B.~T., {Kawaler}, S.~D., {et~al.} 2017, \apjs,
  232, 23

\bibitem[{{Hermes} {et~al.}(2015){Hermes}, {Montgomery}, {Bell}, {Chote},
  {G{\"a}nsicke}, {Kawaler}, {Clemens}, {Dunlap}, {Winget}, \&
  {Armstrong}}]{2015ApJ...810L...5H}
{Hermes}, J.~J., {Montgomery}, M.~H., {Bell}, K.~J., {et~al.} 2015, \apjl, 810,
  L5

\bibitem[{{Jenkins} {et~al.}(2016){Jenkins}, {Twicken}, {McCauliff},
  {Campbell}, {Sanderfer}, {Lung}, {Mansouri-Samani}, {Girouard}, {Tenenbaum},
  {Klaus}, {Smith}, {Caldwell}, {Chacon}, {Henze}, {Heiges}, {Latham},
  {Morgan}, {Swade}, {Rinehart}, \& {Vanderspek}}]{2016SPIE.9913E..3EJ}
{Jenkins}, J.~M., {Twicken}, J.~D., {McCauliff}, S., {et~al.} 2016, in
  \procspie, Vol. 9913, Software and Cyberinfrastructure for Astronomy IV,
  99133E

\bibitem[{{Kalup} {et~al.}(2021){Kalup}, {. Bogn{\'a}r}, \&
  {S{\'o}dor}}]{2021AcA....71..281K}
{Kalup}, C., {. Bogn{\'a}r}, Z., \& {S{\'o}dor}, {\'A}. 2021, \actaa, 71, 281

\bibitem[{{Kim}(2007)}]{2007PhDT........13K}
{Kim}, A. 2007, PhD thesis, The University of Texas at Austin

\bibitem[{{Montgomery} {et~al.}(2020){Montgomery}, {Hermes}, {Winget},
  {Dunlap}, \& {Bell}}]{2020ApJ...890...11M}
{Montgomery}, M.~H., {Hermes}, J.~J., {Winget}, D.~E., {Dunlap}, B.~H., \&
  {Bell}, K.~J. 2020, \apj, 890, 11

\bibitem[{{Mukadam} {et~al.}(2004){Mukadam}, {Mullally}, {Nather}, {Winget},
  {von Hippel}, {Kleinman}, {Nitta}, {Krzesi{\'n}ski}, {Kepler}, {Kanaan},
  {Koester}, {Sullivan}, {Homeier}, {Thompson}, {Reaves}, {Cotter},
  {Slaughter}, \& {Brinkmann}}]{2004ApJ...607..982M}
{Mukadam}, A.~S., {Mullally}, F., {Nather}, R.~E., {et~al.} 2004, \apj, 607,
  982

\bibitem[{{Nather} {et~al.}(1990){Nather}, {Winget}, {Clemens}, {Hansen}, \&
  {Hine}}]{1990ApJ...361..309N}
{Nather}, R.~E., {Winget}, D.~E., {Clemens}, J.~C., {Hansen}, C.~J., \& {Hine},
  B.~P. 1990, \apj, 361, 309

\bibitem[{{Paxton} {et~al.}(2011){Paxton}, {Bildsten}, {Dotter}, {Herwig},
  {Lesaffre}, \& {Timmes}}]{2011ApJS..192....3P}
{Paxton}, B., {Bildsten}, L., {Dotter}, A., {et~al.} 2011, \apjs, 192, 3

\bibitem[{{Ricker} {et~al.}(2015){Ricker}, {Winn}, {Vanderspek}, {Latham},
  {Bakos}, {Bean}, {Berta-Thompson}, {Brown}, {Buchhave}, {Butler}, {Butler},
  {Chaplin}, {Charbonneau}, {Christensen-Dalsgaard}, {Clampin}, {Deming},
  {Doty}, {De Lee}, {Dressing}, {Dunham}, {Endl}, {Fressin}, {Ge}, {Henning},
  {Holman}, {Howard}, {Ida}, {Jenkins}, {Jernigan}, {Johnson}, {Kaltenegger},
  {Kawai}, {Kjeldsen}, {Laughlin}, {Levine}, {Lin}, {Lissauer}, {MacQueen},
  {Marcy}, {McCullough}, {Morton}, {Narita}, {Paegert}, {Palle}, {Pepe},
  {Pepper}, {Quirrenbach}, {Rinehart}, {Sasselov}, {Sato}, {Seager},
  {Sozzetti}, {Stassun}, {Sullivan}, {Szentgyorgyi}, {Torres}, {Udry}, \&
  {Villasenor}}]{2015JATIS...1a4003R}
{Ricker}, G.~R., {Winn}, J.~N., {Vanderspek}, R., {et~al.} 2015, Journal of
  Astronomical Telescopes, Instruments, and Systems, 1, 014003

\bibitem[{{Romero} {et~al.}(2023){Romero}, {da Rosa}, {Kepler}, {Bradley},
  {Uzundag}, {Bell}, {Hermes}, \& {Lauffer}}]{2023MNRAS.518.1448R}
{Romero}, A.~D., {da Rosa}, G.~O., {Kepler}, S.~O., {et~al.} 2023, \mnras, 518,
  1448

\bibitem[{{Romero} {et~al.}(2022){Romero}, {Kepler}, {Hermes}, {Amaral},
  {Uzundag}, {Bogn{\'a}r}, {Bell}, {VanWyngarden}, {Baran}, {Pelisoli},
  {Oliveira}, {Koester}, {Klippel}, {Fraga}, {Bradley}, {Vu{\v{c}}kovi{\'c}},
  {Heintz}, {Reding}, {Kaiser}, \& {Charpinet}}]{2022MNRAS.511.1574R}
{Romero}, A.~D., {Kepler}, S.~O., {Hermes}, J.~J., {et~al.} 2022, \mnras, 511,
  1574

\bibitem[{{Sodor}(2012)}]{2012KOTN...15....1S}
{Sodor}, A. 2012, Konkoly Observatory Occasional Technical Notes, 15, 1

\bibitem[{{Stobie} {et~al.}(1993){Stobie}, {Chen}, {O'Donoghue}, \&
  {Kilkenny}}]{1993MNRAS.263L..13S}
{Stobie}, R.~S., {Chen}, A., {O'Donoghue}, D., \& {Kilkenny}, D. 1993, \mnras,
  263, L13

\bibitem[{{Tremblay} {et~al.}(2011){Tremblay}, {Bergeron}, \&
  {Gianninas}}]{2011ApJ...730..128T}
{Tremblay}, P.~E., {Bergeron}, P., \& {Gianninas}, A. 2011, \apj, 730, 128

\bibitem[{{Tremblay} {et~al.}(2015){Tremblay}, {Ludwig}, {Freytag}, {Fontaine},
  {Steffen}, \& {Brassard}}]{2015ApJ...799..142T}
{Tremblay}, P.-E., {Ludwig}, H.-G., {Freytag}, B., {et~al.} 2015, \apj, 799,
  142

\bibitem[{{Tremblay} {et~al.}(2013){Tremblay}, {Ludwig}, {Steffen}, \&
  {Freytag}}]{2013A&A...559A.104T}
{Tremblay}, P.~E., {Ludwig}, H.~G., {Steffen}, M., \& {Freytag}, B. 2013, \aap,
  559, A104

\bibitem[{{Unno} {et~al.}(1989){Unno}, {Osaki}, {Ando}, {Saio}, \&
  {Shibahashi}}]{1989nos..book.....U}
{Unno}, W., {Osaki}, Y., {Ando}, H., {Saio}, H., \& {Shibahashi}, H. 1989,
  {Nonradial oscillations of stars}

\bibitem[{{Winget} \& {Kepler}(2008)}]{2008ARA&A..46..157W}
{Winget}, D.~E. \& {Kepler}, S.~O. 2008, \araa, 46, 157

\bibitem[{{Winget} {et~al.}(1982){Winget}, {van Horn}, {Tassoul}, {Fontaine},
  {Hansen}, \& {Carroll}}]{1982ApJ...252L..65W}
{Winget}, D.~E., {van Horn}, H.~M., {Tassoul}, M., {et~al.} 1982, \apjl, 252,
  L65

\bibitem[{{Yeates} {et~al.}(2005){Yeates}, {Clemens}, {Thompson}, \&
  {Mullally}}]{2005ApJ...635.1239Y}
{Yeates}, C.~M., {Clemens}, J.~C., {Thompson}, S.~E., \& {Mullally}, F. 2005,
  \apj, 635, 1239

\bibitem[{{Zacharias} {et~al.}(2012){Zacharias}, {Finch}, {Girard}, {Henden},
  {Bartlett}, {Monet}, \& {Zacharias}}]{2012yCat.1322....0Z}
{Zacharias}, N., {Finch}, C.~T., {Girard}, T.~M., {et~al.} 2012, VizieR Online
  Data Catalog, I/322A

\bibitem[{{Zong} {et~al.}(2016){Zong}, {Charpinet}, {Vauclair}, {Giammichele},
  \& {Van Grootel}}]{2016A&A...585A..22Z}
{Zong}, W., {Charpinet}, S., {Vauclair}, G., {Giammichele}, N., \& {Van
  Grootel}, V. 2016, \aap, 585, A22

\end{thebibliography}

\end{document}